\documentclass[12pt]{article}
\usepackage{psfig}

\def\lesssim{\mathrel{\mathpalette\vereq<}}
\def\gtrsim{\mathrel{\mathpalette\vereq>}}
\makeatletter
\def\vereq#1#2{\lower3pt\vbox{\baselineskip1.5pt \lineskip1.5pt
\ialign{$\m@th#1\hfill##\hfil$\crcr#2\crcr\sim\crcr}}}
\makeatother

\begin{document}
\begin{titlepage}
\begin{center}
\hfill    CERN-TH-2000-168\\
~{} \hfill hep-ph/0006157\\

\vskip .3in

{\large \bf The Oscillation Probability of GeV Solar Neutrinos of All Active 
Species}

\vskip 0.5in

Andr\'e de Gouv\^ea

\vskip 0.1in

{\em CERN - Theory Division\\
     CH-1211 Geneva 23, Switzerland}

\end{center}

\vskip .2in

\begin{abstract}
In this paper, I address the oscillation probability of $O$(GeV) neutrinos of
all active flavours produced inside the Sun and detected at the Earth. 
Flavours other than electron-type neutrinos may be produced, for
example, by the annihilation of WIMPs which may be trapped inside the Sun. 
In the GeV energy regime, matter effects are important both for the ``1--3''
system and the ``1--2'' system, and for different neutrino mass hierarchies. 
A numerical scan of the multidimensional three-flavour parameter space is 
performed, ``inspired'' by the current experimental situation. One important
result is that, in the three-flavour oscillation case, 
$P_{\alpha\beta}\neq P_{\beta\alpha}$ for a significant portion of the parameter
space, even if there is no $CP$-violating phase in the MNS matrix. 
Furthermore, $P_{\mu\mu}$ has a significantly different behaviour from 
$P_{\tau\tau}$, which may affect expectations for the number of events detected at 
large neutrino telescopes.  
\end{abstract}

\end{titlepage}

\newpage
\setcounter{footnote}{0}
\setcounter{equation}{0}
\section{Introduction}

In the Standard Model of particle physics, neutrinos are strictly massless. Any
evidence for neutrino masses would, therefore, imply physics beyond the 
Standard Model. Even though the direct experimental determination of a neutrino mass is 
(probably) far beyond the current experimental reach, experiments have been able to 
obtain indirect, and recently very strong, evidence for neutrino masses, 
via neutrino oscillations. 

The key evidence for neutrino oscillations comes from the angular dependent flux of
atmospheric muon-type neutrinos measured at SuperKamiokande \cite{atmospheric}, 
combined with a large deviation of the muon-type to electron-type neutrino flux ratio
from theoretical predictions. This ``atmospheric neutrino puzzle'' is best solved by
assuming that $\nu_{\mu}$ oscillates into $\nu_{\tau}$ and that the $\nu_e$ does not 
oscillate. For a recent analysis of all the atmospheric neutrino data
see \cite{atmos_analysis}. 

On the other hand, measurements of the solar neutrino flux 
\cite{Cl,Kamiokande,GALLEX,SAGE,Super-K} have always been plagued by a large
suppression of the measured solar $\nu_e$ flux with respect to theoretical 
predictions \cite{SSM}. 
Again, this ``solar neutrino puzzle'' is best resolved by assuming that $\nu_e$ oscillates
into a linear combination of the other flavour eigenstates 
\cite{bksreview,rate_analysis} (for a more conservative analysis of the event rates and
the inclusion of the ``dark side'' of the parameter space, see \cite{dark_side}). 
The most recent analysis of the solar neutrino data which includes the mixing of 
three active neutrino species can be found in \cite{solar_3}.

Neutrino oscillations were first hypothesised by Bruno Pontecorvo in the 1950's 
\cite{Pontecorvo}. The hypothesis of three flavour mixing was first raised by 
Maki, Nakagawa and Sakata \cite{MNS}. In light of the solar neutrino
puzzle, Wolfenstein \cite{W} and Mikheyev and Smirnov \cite{MS} realized that 
neutrino--matter interactions could affect in very radical ways the survival probability
of electron-type neutrinos which are produced in the solar core and detected at the 
Earth (MSW effect). 
  
Since then, significant effort has been devoted to understanding the oscillation
probabilities of electron-type neutrinos produced in the Sun. For example, in 
\cite{KP_3} the survival probability of solar electron-type neutrinos was discussed
in the context of three-neutrino mixing including matter effects, and solutions to
the solar neutrino puzzle in this context were studied (for example, in 
\cite{KP_3,MS_3,solar_3}). 

In this paper, the understanding of solar neutrino oscillations is extend to the
case of other active neutrino species ($\nu_{\mu}$, $\nu_{\tau}$, and antineutrinos) 
produced in the solar core. Even though only electron-type neutrinos are produced
by the nuclear reactions which take place in the Sun's innards, it is well know that,
in a number of dark matter models, dark matter particles can be trapped 
gravitationally inside the Sun, and that the annihilation of these should yield a 
flux of high energy neutrinos ($E_{\nu}\gtrsim 1$~GeV)
of all species which may be detectable at the Earth \cite{DM_review}. Indeed,
this is one of the goals of very large ``neutrino telescopes,'' such as AMANDA 
\cite{Amanda} or BAIKAL \cite{Baikal}. It is important to understand how
neutrino oscillations will affect the expected event rates at these 
experiments.\footnote{Some effects have already been studied, in the two-neutrino 
case, in \cite{EFM}.}  

The oscillation probability of all neutrino species has, of course, been studied 
in different contexts, such as in the case of neutrinos produced in the core
of supernovae \cite{supernova} or in the case of neutrinos propagating in constant
electron number densities \cite{barger_etal}. The latter case has been receiving a 
considerable amount of attention from neutrino factory studies \cite{nufact}. 
The case at hand (GeV solar neutrinos) differs significantly from these mentioned 
above, in at least a few of the following: source-detector distance, electron number 
density average value and position dependency, energy average value and spectrum. 
Neutrino factory studies, for example, are interested in $O$(1000)~km base-lines, 
$O$(10)~GeV electron-type and muon-type neutrinos produced via muon decay
propagating in roughly constant, Earth-like (matter densities around 3 g/cm$^3$) 
electron number densities.     

The paper is organised as follows. In Sec.~2, the well known case of two-flavour
oscillations is reviewed in some detail, while special attention will be paid to 
neutrinos produced inside the Sun. In Sec.~3 the same discussion is extended to the less
familiar case of three-flavour oscillations. Again, special attention is paid to 
neutrinos produced in the Sun's core. In Sec.~4 the results presented in Sec.~3 will
be analysed numerically, and the three-neutrino multi-dimensional 
parameter space will be explored. Sec.~5 contains a summary of the results and the
conclusions.

It is important to comment at this point that one of the big challenges of studying
three-flavour oscillations is the multi-dimensional parameter space, composed of
three mixing angles, two mass-squared differences, and one complex phase, plus the
neutrino energy. For this reason, the discussions presented here will take advantage
of the current experimental situation to constrain the parameter space, and of the
possibility of producing neutrinos of all species via dark matter annihilations to 
constrain the neutrino energies to the range from a few to tens of GeV.    

\setcounter{equation}{0}
\section{Two-Flavour Oscillations}

In this section, the well studied case of two-flavour oscillations will
be reviewed \cite{general_review}. 
This is done in order to present the formalism which will be later
extended to the case of three-flavour oscillations and describe general 
properties of neutrino oscillations and of neutrinos produced in the Sun's core.

\subsection{Generalities}

Neutrino oscillations take place because, similar to what happens in the quark
sector, neutrino weak eigenstates are different from neutrino mass eigenstates.
The two sets are related by a unitary matrix, which is, in the case of two-flavour 
mixing, parametrised
by one mixing angle $\vartheta$.\footnote{If the neutrinos are Majorana particles, 
there is also a Majorana phase, which will be ignored throughout since it plays 
no role in the physics of neutrino oscillations.}
\begin{equation}
\left(\matrix{\nu_{e} \cr \nu_{x} }\right)=
\left(\matrix{U_{e1}&U_{e2}\cr
U_{x1}&U_{x2}}\right) \left(\matrix{\nu_{1} \cr \nu_{2} }\right)=
\left(\matrix{\cos\vartheta&\sin\vartheta\cr
-\sin\vartheta&\cos\vartheta}\right) \left(\matrix{\nu_{1} \cr \nu_{2} }\right),
\end{equation} 
where $\nu_1$ and $\nu_2$ are neutrino mass eigenstates with masses $m_1$ and 
$m_2$, respectively, and $\nu_x$ is the flavour eigenstate orthogonal to $\nu_e$.
All physically distinguishable situations can be obtained if 
$0\leq\vartheta\leq\pi/2$ and 
$m_1^2\leq m_2^2$ or $0\leq\vartheta\leq\pi/4$ and no constraint is 
imposed on the masses-squared.

In the case of oscillations in vacuum, it is trivial to compute the probability
that a neutrino produced in a flavour state $\alpha$ is detected as a neutrino of 
flavour $\beta$, assuming that the neutrinos are ultrarelativistic and propagate
with energy $E_{\nu}$:
\begin{equation}
P_{\alpha\beta}=|U_{\beta1}|^2|U_{\alpha1}|^2+|U_{\beta2}|^2|U_{\alpha2}|^2+2Re\left(
U_{\beta1}^*U_{\beta2}U_{\alpha1}U_{\alpha2}^*
e^{i\frac{\Delta m^2x}{2E_{\nu}}}\right).
\end{equation}
Here $\Delta m^2\equiv m^2_{2}-m_1^2$ is the mass-squared difference between
the two mass eigenstates and $x$ is the distance from the detector to the source.
It is trivial to note that $P_{\alpha\beta}=P_{\beta\alpha}$ 
since all $U_{\alpha i}$ are real and the
theory is $T$-conserving. Furthermore, note that $\vartheta$ is indistinguishable from
$\pi/2-\vartheta$ (or, equivalently, the sign of $\Delta m^2$ is not physical), 
and all physically distinguishable situations are obtained by allowing 
$0\leq\vartheta\leq\pi/4$
and choosing a fixed sign for $\Delta m^2$.

In the case of nontrivial neutrino--medium interactions, the computation of
$P_{\alpha\beta}$ can be rather involved. Assuming that the neutrino--medium 
interactions can be expressed in terms of an effective potential for the neutrino
propagation, one has to solve
\begin{equation}
\frac{\rm d}{{\rm d}t}\left(\matrix{\nu_{1}(t) \cr \nu_{2}(t) }\right)=
-i\left[\left(\matrix{E_1 & 0 \cr
0 & E_2}\right)+\left(\matrix{V_{11}(t) & V_{12}(t) \cr
V_{12}(t)^* & V_{22}(t)}\right)\right]\left(\matrix{\nu_{1}(t) \cr 
\nu_{2}(t) }\right),
\end{equation}
with the appropriate boundary conditions (either a $\nu_e$ or a $\nu_x$ as the 
initial state, for example). In the ultrarelativistic limit one may approximate
$E_2-E_1\simeq \Delta m^2/2E_{\nu}$, ${\rm d}/{\rm d}t=\simeq{\rm d}/{\rm d}x$, and 
$V_{ij}(t)\simeq V_{ij}(x)$. A very crucial assumption is that there is no 
kind of neutrino absorption due to the neutrino--medium interaction, {\it i.e.,}\/
the $2\times 2$ Hamiltonian for the neutrino system is Hermitian. 

It is interesting to argue what can be said about
$P_{\alpha\beta}$ in very general terms. First, the conservation of probability 
requires
that 
\begin{eqnarray}
P_{ee}+P_{ex}&=&1, \\
P_{xe}+P_{xx}&=&1.
\end{eqnarray}
Second, given that the Hamiltonian evolution is unitary, 
\begin{equation}
P_{ee}+P_{xe}=1.
\label{extra_constraint}
\end{equation}
It is easy to show that the extra constraint $P_{ex}+P_{xx}=1$ is redundant.
Eq.~(\ref{extra_constraint}) 
can be understood by the following ``intuitive'' argument: if the same
amount of $\nu_e$ and $\nu_x$ is produced, independent of what happens to $\nu_e$ and
$\nu_x$ during flight, the number of $\nu_e$ and $\nu_x$ detected in the end
has to be the same. In light of the constraints above, one can show that there is
only one independent $P_{\alpha\beta}$, which is normally chosen to be $P_{ee}$. 
The others are 
given by $P_{ex}=P_{xe}=1-P_{ee}$ and $P_{xx}=P_{ee}$. Note that the equality
$P_{ex}=P_{xe}$ is {\it not}\/ a consequence of $T$-invariance, but a consequence
of the unitarity of the Hamiltonian evolution and particular only to the 
two-flavour oscillation case, as will be shown later.  

\subsection{Oscillation of Neutrinos Produced in the Sun's Core}

It is well known \cite{W,MS}
that neutrino--Sun interactions affect the oscillation probabilities
of neutrinos produced in the Sun's core in very nontrivial ways. Indeed, all but one
solution to the solar neutrino puzzle rely heavily on neutrino--Sun interactions
\cite{bksreview,rate_analysis,dark_side}. 
The survival probability of electron-type solar neutrinos has been 
computed in many different approximations by a number of people over the years, and
can be understood in very simple terms \cite{general_review}. 

In the presence of electrons, the differential equation satisfied by the two neutrino
system is, in the flavour basis,
\begin{equation}
\frac{\rm d}{{\rm d}x}\left(\matrix{\nu_{e}(x) \cr \nu_{x}(x) }\right)=-i\left[
\frac{\Delta m^2}{2E_{\nu}}\left(\matrix{|U_{e2}|^2 & U_{e2}^*U_{\mu2} \cr
U_{e2}U_{\mu2}^* & |U_{\mu2}|^2 }\right)
+\left(\matrix{A(x) & 0 \cr 0 & 0}\right)\right]
\left(\matrix{\nu_{e}(x) \cr \nu_{x}(x) }\right),
\label{eq_2ns}
\end{equation} 
where terms proportional to the $2\times 2$ identity matrix were neglected, since they
play no role in the physics of neutrino oscillations. 
\begin{equation}
A(x)=\sqrt{2}G_FN_e(x)
\label{A(x)}
\end{equation} 
is the charged current contribution to the $\nu_e$-$e$ forward scattering amplitude, 
$G_F$ is Fermi's constant, and $N_e(x)$ is the position dependent electron number
density. In the case of the Sun \cite{SSM} (see also \cite{bahcall_www}), 
$A\equiv A(0)\simeq 6\times 10^{-3}$~eV$^2$/GeV, 
assuming an average core density of 79~g/cm$^3$, and $A(x)$ falls roughly 
exponentially until close to the Sun's edge. It is safe to say that significantly
far away from the Sun's edge $A(x)$ is zero.

A particularly simple way of understanding the propagation of electron-type neutrinos
produced in the Sun's core to the Earth is to start with a $\nu_e$ state in the basis 
of the eigenstates of the Hamiltonian evaluated at the production point, 
$|\nu_e\rangle=cos\vartheta_M(0)|\nu_L\rangle+\sin\vartheta_M(0)|\nu_H\rangle$, 
where $|\nu_H\rangle$
($|\nu_L\rangle$) correspond to the highest (lowest) instantaneous Hamiltonian 
eigenstate. The matter mixing angle $\vartheta_M\equiv\vartheta_M(0)$ is given by
\begin{equation}
\cos 2\vartheta_M=\frac{\Delta m^2\cos 2\vartheta-2E_{\nu}A}
{\sqrt{(\Delta m^2)^2+A^2-4E_{\nu}A\Delta m^2\cos 2\vartheta}}.
\label{cos2tm}
\end{equation}

The evolution of this initial state from the Sun's core is described by an arbitrary 
unitary matrix until the neutrino reaches the Sun's edge. From this point on, one can 
rotate the state to the mass basis and follow the vacuum evolution of the state. 
Therefore, $P_{ee}(x)$, where $x$ is is the distance from the Sun's edge to some
point far away from the Sun (for example, the Earth), is
\begin{equation}
P_{ee}(x)=\left|\left(\matrix{U_{e1}^* & U_{e2}^*}\right)\left(\matrix{1 & 0 \cr
0 & e^{-i\frac{\Delta m^2x}{2E_{\nu}}}}\right)\left(\matrix{A & B \cr
-B^* & A^*}\right)
\left(\matrix{\cos\vartheta_M \cr \sin\vartheta_M}\right)\right|^2,
\label{Peex}
\end{equation}  
where overall phases in the amplitude have been neglected.
The matrix parametrised by $A,B$ 
represents the evolution of the system from the 
Sun's core to vacuum, and also rotates the state into the mass basis.\footnote{The
most general form of a $2\times 2$ unitary matrix is $\left(\matrix{A & B \cr
-B^* & A^*}\right)\left(\matrix{1 & 0 \cr 0 & e^{i\zeta}} \right)$, where 
$|A|^2+|B|^2=1$ and $0\leq\zeta\leq 2\pi$. In the case of neutrino oscillations,
however, the physical quantities are $|A|^2$ and the phase of $AB^*$, and therefore
$\zeta$ can be ignored.}
Expanding Eq.~(\ref{Peex}), and assuming that there is no coherence in the Sun's core
between $\nu_L$ and $\nu_H$,\footnote{This is in general the case, because one
has to consider that neutrinos are produced at different points in space and time.}
one arrives at the well known expression (these have been first derived using a different 
language in \cite{Petcov_eq} and \cite{PP})
\begin{equation}
P_{ee}(x)=P_1\cos^2\vartheta + P_2\sin^2\vartheta -\cos 2\vartheta_M
\sqrt{P_c(1-P_c)}\sin 2\vartheta\cos\left(\frac{\Delta m^2x}{2E_{\nu}}+\delta
\right),
\label{pee}
\end{equation}
where $\delta$ is the phase of $AB^*$, $P_c\equiv |B|^2=1-|A|^2$ is the ``level crossing
probability'', and 
$P_1=1-P_2=\frac{1}{2}+\frac{1}{2}\left(1-2P_c\right)\cos 2\vartheta_M$ is 
interpreted as the probability that the neutrino exits the Sun as a 
$\nu_1$. 

Eq.~(\ref{pee}) should be valid in all cases of interest, 
and contains a large amount of
features. In the case of the solar neutrino puzzle, the neutrino energies of interest
range between hundreds of keV to ten MeV, and matter effects start to play a role for
values of $\Delta m^2$ as high as $10^{-4}$~eV$^2$. In the adiabatic limit 
($P_c\rightarrow 0$) very small values of $P_{ee}$ are attainable when 
$\cos 2\vartheta_M\rightarrow -1$ and $\sin^2\vartheta$ is small. More generally, 
in this
limit $P_{ee}=\sin^2\vartheta$. This is what happens for all solar neutrino energies
in the case of the LOW solution,\footnote{See 
\cite{bksreview,rate_analysis,dark_side} for the labelling of the regions of the 
parameter space that solve the solar neutrino puzzle} 
for solar neutrino energies above a few MeV 
in the case of the 
LMA solution, and for 400~keV~$\lesssim E_{\nu}\lesssim 1$~MeV energies in the case 
of the SMA solution. In the extremely nonadiabatic limit,
which is reached when $\Delta m^2/2E_{\nu}\ll A$, 
$P_c\rightarrow \cos^2\vartheta$ and
$\cos 2\vartheta_M\rightarrow -1$, the original vacuum oscillation expression is 
obtained,
up to the ``matter phase'' $\delta$. This is generically what happens in the VAC 
solution to the solar neutrino puzzle.

If the electron number density is in fact exponential, one can solve 
Eq.~(\ref{eq_2ns}) exactly \cite{exponential,PC}. 
For $N_e(x)=N_e(0)~e^{-x/r_0}$, where $x=0$ is the centre of the Sun,
\begin{equation}
P_c=\frac{e^{-\gamma\sin^2\vartheta}-e^{-\gamma}}{1-e^{-\gamma}},
\label{pc} 
\end{equation}  
\cite{PC,check} where 
\begin{equation}
\gamma=2\pi r_0\frac{\Delta m^2}{2E_{\nu}}=1.05\left(\frac{\Delta m^2}
{10^{-6}~{\rm eV}^2}\right)\left(\frac{1~{\rm GeV}}{E_{\nu}}\right),
\label{gamma}
\end{equation}
for $r_0=R_{\odot}/10.54=6.60\times 10^4$~km \cite{bahcall_www}. In the case of the Sun, 
the exponential profile approximation has been examined \cite{check}, and was
shown to be very accurate, especially if one allows $r_0$ to vary as a 
function of $\Delta m^2/2E_{\nu}$.

The exact expression for $\delta$ has also been obtained \cite{Petcov_eq}, 
and the readers are referred to \cite{P_phase} 
for details concerning physical implications of the matter phase. Its effects
will not be discussed here any further. 

\subsection{The Case of Antineutrinos}

Antineutrinos that are produced in the Sun's core obey a differential equation
similar to Eq.~(\ref{eq_2ns}), except that the sign of the matter potential 
changes, {\it i.e.} $A(x)\leftrightarrow -A(x)$, and 
$U_{\alpha i}\leftrightarrow U^*_{\alpha i}$ 
(this is immaterial since, in the two-flavour mixing case, all $U_{\alpha i}$ 
are real). 

Instead of working out the probability of an electron-type antineutrino being
detected as an electron-type antineutrino $P_{\bar{e}\bar{e}}$ from scratch, 
there is a very simple way
of relating it to $P_{ee}$. One only has to note that, if the following transformation
is applied to Eq.~(\ref{eq_2ns}): $\vartheta\rightarrow \pi/2-\vartheta$, subtract the
matrix $A(1_{2\times 2})$, where $1_{2\times 2}$ is the $2\times 2$ identity matrix
and relabel $\nu_e(x)\leftrightarrow \nu_x(x)$, the equation of motion for 
antineutrinos is obtained.\footnote{If one decides to limit 
$0\leq\vartheta\leq\pi/4$, 
a similar result can be obtained if $\Delta m^2\rightarrow-\Delta m^2$, 
explicitly $P_{\bar{e}\bar{e}}(\Delta m^2)=P_{ee}(-\Delta m^2)$.} 
Therefore, $P_{\bar{e}\bar{e}}(\vartheta)=
P_{xx}(\pi/2-\vartheta)=P_{ee}(\pi/2-\vartheta)$ (this was pointed out in 
\cite{Chizhov}). Remember that, in the case of vacuum
oscillations, $\vartheta$ is physically equivalent to $\pi/2-\vartheta$, so 
$P_{\bar{e}\bar{e}}=P_{ee}$. In the more general case of nontrivial matter effects,
this is clearly not the case, since the presence of matter (or antimatter) explicitly
breaks $CP$-invariance. 

It is curious to note that, in the case of two-flavour oscillations, there is no
$T$-noninvariance, {\it i.e.,}\/ $P_{\alpha\beta}=P_{\beta\alpha}$, while there is 
potentially large $CP$ violation, {\it i.e.,}\/ $P_{\alpha\beta}\neq 
P_{\bar{\alpha}\bar{\beta}}$, even if the Hamiltonian for the system is explicitly
$T$-noninvariant and $CP$-noninvariant, as is the case of the propagation of
neutrinos produced in the Sun (namely $A(t)$ is a generic function of time
and $A(t)$ for neutrinos is $-A(t)$ for antineutrinos). 

\setcounter{footnote}{0}
\setcounter{equation}{0}
\section{Three Flavour Oscillations}

Currently, aside from the solar neutrino puzzle, there is an even more convincing
evidence for neutrino oscillations, namely the suppression of the muon-type
neutrino flux in atmospheric neutrino experiments \cite{atmospheric}. 
This atmospheric neutrino puzzle
is best solved by $\nu_{\mu}\leftrightarrow\nu_{\tau}$ oscillations with a large
mixing angle \cite{atmos_analysis}. 
Furthermore, the values of $\Delta m^2$ required to solve the atmospheric
neutrino puzzle are at least one order of magnitude higher than the values required 
to solve
the solar neutrino puzzle. For this reason, in order to solve both neutrino puzzles
in terms of neutrino oscillations, three neutrino families are required.

In this section, the oscillations of three neutrino flavours will considered. 
In order to simplify the discussion, I will concentrate on neutrinos
with energies ranging from a few to tens of GeV, which is the energy range
expected for neutrinos produced by the annihilation of 
dark matter particles which are possibly trapped
inside the Sun. Furthermore, a number of experimentally inspired 
constraints on the neutrino
oscillation parameter space will be imposed, as will become clear later.  

\subsection{Generalities}

Similar to the two-flavour case, the ``mapping'' between the flavour eigenstates,
$\nu_e$, $\nu_{\mu}$ and $\nu_{\tau}$ and the mass eigenstates $\nu_i$, $i=1,2,3$ with
masses $m_i$ can be performed with a general $3\times 3$ unitary matrix, which is
parametrised by three mixing angles ($\theta$, $\omega$, and $\xi$) and a complex
phase $\phi$. In short hand notation $\nu_{\alpha}=U_{\alpha i}\nu_{i}$ where 
$\alpha=e,\mu,\tau$ and $i=1,2,3$. The MNS mixing matrix \cite{MNS} 
will be written, similar
to the standard CKM quark mixing matrix \cite{PDG}, as {\small
\begin{equation}
\left(\matrix{U_{e1} & 
U_{e2} & U_{e3} \cr U_{\mu1} & U_{\mu2} & U_{\mu3} \cr 
U_{\tau1} & U_{\tau2} & U_{\tau3}}\right)
=\left(\matrix{c\omega~c\xi & 
s\omega~c\xi & s\xi e^{i\phi} \cr -s\omega~c\theta-c\omega~s\theta~s\xi e^{-i\phi} 
& c\omega~c\theta- s\omega~s\theta~s\xi e^{-i\phi} & s\theta~c\xi \cr 
s\omega~s\theta-c\omega~c\theta~s\xi e^{-i\phi}
& -c\omega~s\theta-s\omega~c\theta~s\xi e^{-i\phi} 
& c\theta~c\xi}\right),
\label{MNSmatrix}
\end{equation} }
where $c\zeta\equiv\cos\zeta$ and $s\zeta\equiv\sin\zeta$ for 
$\zeta=\omega,\theta,\xi$. If the neutrinos are Majorana particles, two extra phases
should be added to the MNS matrix, but, since they play no role in the physics 
of neutrino oscillations, they can be safely ignored.
All physically distinguishable situations can be 
obtained if one allows $0\leq\phi\leq\pi$, all
angles to vary between $0$ and $\pi/2$ and no
restriction is imposed on the sign of the mass-squared differences, 
$\Delta m^2_{ij}\equiv m^2_i-m^2_j$. Note that there are only two independent 
mass-squared differences, which are chosen here to be $\Delta m^2_{21}$ and 
$\Delta m^2_{31}$.

All experimental evidence from solar, atmospheric, and reactor neutrino experiments
\cite{atmospheric,Cl,Kamiokande, GALLEX,SAGE,Super-K,reactor}
can be satisfied,\footnote{There is 
evidence for neutrino oscillations coming from the LSND experiment \cite{LSND}. 
Such evidence 
has not yet been confirmed by another experiment, and will not be considered
in this paper.
If, however, it is indeed confirmed, it is quite likely that a fourth, sterile, 
neutrino will have to be introduced into the picture.} somewhat conservatively,  
by assuming \cite{general_review}: $10^{-4}$~eV$^2\lesssim
|\Delta m^2_{31}|\simeq|\Delta m^2_{32}|\lesssim 10^{-2}$~eV$^2$, 
$0.3\lesssim\sin^2\theta\lesssim 0.7$, $10^{-11}$~eV$^2\lesssim
|\Delta m^2_{21}|\lesssim 10^{-4}$~eV$^2$, 
$\sin^2\xi\lesssim 0.1$, while
$\omega$ is mostly unconstrained. There is presently no information on $\phi$. In
determining these bounds, it was explicitly assumed that only three active neutrinos
exist.

A few comments about the constraints imposed above are in order. First, one may 
complain that $\omega$ is more constrained than mentioned above by the solar neutrino
data. The situation is far from definitive, however. As pointed out recently in 
\cite{dark_side} if the uncertainty on the $^8$B neutrino flux is inflated or if some
of the experimental data is not considered (especially the Homestake 
data \cite{Cl}) in the fit, a
much larger range of $\Delta m^2_{21}$ and $\omega$ is allowed. Furthermore, 
if three-flavour mixing is considered \cite{solar_3}, different regions in
the parameter space $\Delta m^2_{21}$-$\sin^2\omega$ are allowed for different values
of $\sin^2\xi$, even if $\sin^2\xi$ is constrained to be small.

Second, the limit from the Chooz and Palo Verde reactor experiments \cite{reactor}
do not constrain $\sin^2\xi$ for $|\Delta m^2_{31}|\lesssim 10^{-3}$~eV$^2$. 
Furthermore, their constraints
are to $\sin^2 2\xi$, so values of $\sin^2\xi$ close to one should also be allowed. 
However, the constraints from the atmospheric neutrino data require $\cos^2\xi$ to be
close to one. This is easy to understand. Assuming that $L_{21}^{\rm osc}$ is much 
larger than the Earth's diameter and that $\Delta m^2_{31}=\Delta m^2_{32}$,
\begin{equation}
P_{\mu\mu}^{\rm atm}=1-4\cos^2\xi\sin^2\theta(1-\cos^2\xi\sin^2\theta)
\sin^2\left(\frac{\Delta m^2_{31}x}{4E_{\nu}}\right),
\end{equation} 
according to upcoming Eq.~(\ref{p3vac}).
Almost maximal mixing implies that $\cos^2\xi\sin^2\theta\simeq 1/2$. With the further
constraint from $P_{ee}^{\rm atm}$, namely $\sin^2 2\xi\simeq 0$, 
one concludes that $\cos^2\xi\simeq 1$ and $\sin^2\theta\simeq 1/2$. 

In the case of oscillations in vacuum, it is straight forward to compute the 
oscillation
probabilities $P_{\alpha\beta}$ of detecting a flavour $\beta$ given that a flavour
$\alpha$ was produced. 
\begin{equation}
\label{p3vac}
P_{\alpha\beta}=\sum_{i,j}U_{\alpha i}^*U_{\alpha j}U_{\beta i}
U_{\beta j}^*e^{i\frac{\Delta m^2_{ij}x}{2E_{\nu}}}
\end{equation}
The three different oscillation lengths, $L_{\rm osc}^{ij}$, are numerically given by
\begin{equation}
L_{\rm osc}^{ij}=
\frac{4\pi E_{\nu}}{\Delta m^2_{ij}}=2.47\times10^{8}{\rm km}\left(\frac{E}
{1~\rm GeV}\right)
\left(\frac{10^{-8}~\rm eV^2}{\Delta m^2_{ij}}\right),
\end{equation}
which are to be compared to the Earth-Sun distance (1 a.u.$=1.496\times 10^{8}$~km). 
In the energy range of interest, 1~Gev$\lesssim E_{\nu}\lesssim 100$~GeV and given
the experimental constraints on the parameter space described above, it is easy to see
that $L_{\rm osc}^{31}$ and $L_{\rm osc}^{32}$ are much smaller than 1 a.u., and that
its effects should ``wash out'' due to any realistic neutrino energy spectrum, 
detector 
energy resolution, or other ``physical'' effects. Such terms will therefore
be neglected henceforth. In contrast, $L_{\rm osc}^{21}$ maybe as large as 
(and maybe even much larger than!) the Earth-Sun distance. Note that a nonzero
phase $\phi$ implies $T$-violation, {\it i.e.,}\/ $P_{\alpha\beta}\neq  
P_{\beta\alpha}$, unless $L_{\rm osc}^{21}\gg 1$~a.u.. This will be discussed in more
detail later.

In the presence of neutrino--medium interactions, the situation is, in general, more
complicated (indeed, much more!). Similar to the two-neutrino case, it is important
to discuss what is known about the oscillation probabilities. From the conservation
of probability one has    
\begin{eqnarray}
P_{ee}+P_{e\mu}+P_{e\tau}&=&1, \nonumber \\
P_{\mu e}+P_{\mu\mu}+P_{\mu\tau}&=&1, \\
P_{\tau e}+P_{\tau\mu}+P_{\tau\tau}&=&1, \nonumber 
\end{eqnarray}
and, similar to the two-neutrino case, unitarity of the Hamiltonian evolution implies
\begin{eqnarray}
P_{ee}+P_{\mu e}+P_{\tau e}&=&1, \nonumber \\
P_{e\mu}+P_{\mu\mu}+P_{\tau\mu}&=&1, 
\label{const3}
\end{eqnarray}
A third equation of this kind, $P_{e\tau}+P_{\mu\tau}+P_{\tau\tau}=1$, is redundant.
As before, Eqs.~(\ref{const3}) can be understood by arguing that, if equal numbers
of all neutrino species are produced, the number of $\nu_{\beta}$'s to be detected
should be the same, regardless of $\beta$, simply because the neutrino propagation
is governed by a unitary operator.
 
One may therefore express all $P_{\alpha\beta}$ in terms of only four quantities. 
Here, these are chosen to be $P_{ee}$, $P_{e\mu}$, $P_{\mu\mu}$, and $P_{\tau\tau}$. 
The others are given by
\begin{eqnarray}
P_{e\tau} & = & 1-P_{ee}-P_{e\mu}, \nonumber  \\
P_{\mu e} & = & 1+P_{\tau\tau}-P_{ee}-P_{\mu\mu}-P_{e\mu}, \nonumber \\
P_{\mu\tau} & = & P_{ee}+P_{e\mu}-P_{\tau\tau}, \\
P_{\tau e} &= & P_{\mu\mu}+P_{e\mu}-P_{\tau\tau}, \nonumber \\
P_{\tau\mu} & = & 1-P_{\mu\mu}-P_{e\mu}. \nonumber 
\end{eqnarray}
Note that, in general, $P_{\alpha\beta}\neq P_{\beta\alpha}$. 

\subsection{Oscillation of Neutrinos Produced in the Sun's Core}

The propagation of neutrinos in the Sun's core can, similar to the two-neutrino
case, be described by the differential equation
\begin{equation}
\frac{\rm d}{{\rm d}x}\nu_{\alpha}(r)=-i\left(
\sum_{i=2}^{3}\left(\frac{\Delta m^2_{i1}}{2E_{\nu}}\right)U_{\alpha i}^*U_{\beta i}
+A(x) \delta_{\alpha e}\delta_{\beta e}\right) \nu_{\beta}(r),
\label{eq_3nus}
\end{equation}
where $\delta_{\eta\zeta}$ is the Kronecker delta symbol. 
Terms proportional to the identity $\delta_{\alpha\beta}$ are neglected because
they play no role in the physics of neutrino oscillations. The matter induced
potential $A(x)$ is given by Eq.(\ref{A(x)}). 

As in the two-neutrino case, it is useful to first discuss the initial states
$\nu_{\alpha}$ in the Sun's core, and to express them in the basis of instantaneous
Hamiltonian eigenstates, which will be referred to as $|\nu_H\rangle$, 
$|\nu_M\rangle$, and $|\nu_L\rangle$ ($H=$~high, $M$=~medium, and $L$=~low). 
Therefore
\begin{equation}
|\nu_{\alpha}\rangle=H_{\alpha}|\nu_H\rangle+M_{\alpha}|\nu_M\rangle+L_{\alpha}
|\nu_L\rangle,
\end{equation}
where $\langle\nu_{\alpha}|\nu_{\alpha '}\rangle=\delta_{\alpha\alpha '}$. As before
(see Eq.~{\ref{Peex}}),
the probability of detecting this initial state as a $\beta$-type neutrino far
away from the Sun ({\it e.g.,}\/ at the Earth) is given by
\begin{equation}
P_{\alpha\beta}=\left|\left(\matrix{U_{\beta1}^* & U_{\beta2}^* & U_{\beta3}^*}\right)
\left(\matrix{1 & 0 & 0 \cr
0 & e^{-i\frac{\Delta m^2_{21}x}{2E_{\nu}}} & 0 \cr 
0 & 0 & e^{-i\frac{\Delta m^2_{31}x}{2E_{\nu}}}}\right)\left(V_{3\times 3}\right)
\left(\matrix{L_{\alpha} \cr M_{\alpha} \cr H_{\alpha}}\right)\right|^2,
\label{palphabetax}
\end{equation} 
where $V_{3\times 3}$ is an arbitrary $3\times 3$ unitary matrix which takes
care of propagating the initial state until the edge of the Sun and rotating
the state into the mass basis.

In order to proceed, it is useful take advantage of the constraints on the neutrino 
parameter space and the energy range of interest. Note that 
$A\gtrsim\frac{|\Delta m^2_{31}|}{2E_{\nu}}\gg 
\frac{|\Delta m^2_{21}|}{2E_{\nu}}$ (remember
that the energy range of interest is 1~GeV$\lesssim E_{\nu}\lesssim 100$~GeV and that
$A\simeq 6\times 10^{-3}$~eV$^2$/GeV). It has been shown explicitly \cite{KP_3}, 
assuming the neutrino mass-squared hierarchy to be $m_3^2>m_2^2>m_1^2$,
\footnote{I will 
work under this assumption for the time being.} 
that, if the mass-squared differences are very hierarchical ($|\Delta m^2_{31}|\gg
|\Delta m^2_{21}|$), 
the three-level system ``decouples'' into
two two-level systems, {\it i.e.,}\/ one can first deal with matter effects 
in the ``$H-M$'' system and then with the matter effects in the ``$M-L$'' system.
One way of understanding why this is the case is to realize that the ``resonance
point'' corresponding to the $\Delta m^2_{31}$ is very far away from the resonance
point corresponding to $\Delta m^2_{21}$. With this in mind, it is fair to approximate
(this is similar to what is done, for example, in \cite{P_3})
\begin{equation}
V_{3\times 3}=\left(\matrix{A^L & B^L & 0 \cr
-B^{L*} & A^{L*} & 0 \cr 0 & 0 & 1} \right)\left(\matrix{1 & 0 & 0  \cr
0 & A^{H} & B^{H} \cr 0 & -B^{H*} & A^{H*}} \right),
\end{equation} 
where $|B^H|^2=1-|A^H|^2\equiv P_c^H$, $|B^L|^2=1-|A^L|^2\equiv P_c^L$. The 
superscripts $H$, $L$ correspond to the ``high'' and the ``low'' resonances, 
respectively.

It also possible to obtain an approximate expression for the initial states
in the Sun's core. Following the result outline above, this state should be
described by two matter angles, $\xi_M$ and $\omega_M$, corresponding to each
of the two-level systems. Both should be given by Eq.~(\ref{cos2tm}), where,
in the case of $\cos 2\xi_M$, $\vartheta$ is to be replaced by $\xi$ and $\Delta
m^2$ by $\Delta m^2_{31}$, while in the case of $\cos 2\omega_M$,
$\vartheta$ is to be replaced by $\omega$, $\Delta m^2$ by $\Delta m^2_{21}$ 
and $A$ is to be replaced by $A\cos\xi$ \cite{P_3,solar_3}.
Furthermore, because $A\cos\xi\gg\frac{|\Delta m^2_{21}|}{2E_{\nu}}$, 
$\cos 2\omega_M$ can be safely replaced by -1 (remember that $\cos^2\xi\gtrsim0.9$).
Within these approximations, in the Sun's core,
\begin{eqnarray}
\label{inistate3}
|\nu_{e}\rangle&=&\sin\xi_M|\nu_H\rangle+\cos\xi_M|\nu_M\rangle, \nonumber \\
|\nu_{\mu}\rangle&=&\sin\theta\cos\xi_M|\nu_H\rangle-
\sin\theta\sin\xi_M|\nu_M\rangle
-\cos\theta|\nu_L\rangle, \\
|\nu_{\tau}\rangle&=&\cos\theta\cos\xi_M|\nu_H\rangle-
\cos\theta\sin\xi_M|\nu_M\rangle+\sin\theta|\nu_L\rangle. \nonumber
\end{eqnarray} 
The accuracy of this approximation has been tested numerically in the range
of parameters of interest, and the difference between the ``exact'' result and
the approximate result presented in Eq.~(\ref{inistate3}) is negligible.

Keeping all this in mind, it is straight forward to compute all oscillation 
probabilities, starting from Eq.~(\ref{palphabetax}). From here on, $\phi=0$ 
(no $T$-violating phase in the mixing matrix, such that all $U_{\alpha i}$ are real)
will be assumed, in order to simplify expressions and render the results cleaner. 
In the end of the day one obtains
\begin{eqnarray}
\label{pall3}
P_{\alpha\beta}&=&a_{\alpha}^2 (U_{\beta 1})^2 + b_{\alpha}^2 (U_{\beta 2})^2 + 
c_{\alpha}^2 (U_{\beta 3})^2 +  2a_{\alpha}b_{\alpha} (U_{\beta1}U_{\beta2}) 
\cos\left(\frac{\Delta m^2_{21}x}{2E_{\nu}}+\delta^L\right) \nonumber \\
&&{\rm or} \\
P_{\alpha\beta}&=&\left(a_{\alpha}U_{\beta 1} + b_{\alpha}U_{\beta 2}\right)^2 + 
c_{\alpha}^2 (U_{\beta 3})^2 - 4a_{\alpha}b_{\alpha} (U_{\beta1}U_{\beta2}) 
\sin^2\left(\frac{\Delta m^2_{21}x}{4E_{\nu}}+\delta^L\right), \nonumber
\end{eqnarray}
where $\delta^L$ is the matter phase, induced in the low resonance, and   
\begin{eqnarray}
a_e&=&\sqrt{P_2^HP_c^L}, \nonumber \\
b_e&=&\sqrt{P_2^H(1-P_c^L)}, \nonumber \\
c_e&=&\sqrt{P_3^H}, \nonumber \\
a_{\mu}&=&-\sqrt{(1-P_c^L)}\cos\theta-\sqrt{P_3^HP_c^L}\sin\theta, 
\nonumber \\
b_{\mu}&=&\sqrt{P_c^L}\cos\theta-\sqrt{P_3^H(1-P_c^L)}\sin\theta, \\
c_{\mu}&=&\sqrt{P_2^H\sin^2\theta}, \nonumber \\
a_{\tau}&=&\sqrt{(1-P_c^L)}\sin\theta-\sqrt{P_3^HP_c^L}\cos\theta, 
\nonumber \\
b_{\tau}&=&-\sqrt{P_c^L}\sin\theta-\sqrt{P_3^H(1-P_c^L)}\cos\theta,
\nonumber \\
c_{\tau}&=&\sqrt{P_2^H\cos^2\theta}, \nonumber 
\end{eqnarray}
and $P_2^H=1-P_3^H=(|A^H|^2\cos^2\xi_M+|B^H|^2\sin^2\xi_M)$, which can also be 
written as $P_2^H=\frac{1}{2}+\frac{1}{2}\left(1-2P_c^H\right)\cos 2\xi_M$. This is
to be compared with the expression for $P_1$ obtained in the two-flavour case. 
Note that $a^2_{\alpha}+b^2_{\alpha}+c^2_{\alpha}=1$. The effect of $\delta^L$ will
not be discussed here and from here on $\delta^L$ will be set to zero. For details
about the significance of $\delta^L$ for solar neutrinos in the two-flavour
case, readers are referred to \cite{Petcov_eq,P_phase}.

Many comments are in order. First, in the nonadiabatic limit which can be obtained for
very large energies, $P_c^H\rightarrow \cos^2\xi$, $P_c^L\rightarrow 
\cos^2\omega$ and
$\cos 2\xi_M\rightarrow -1$. It is trivial to check that in this limit 
$a_{\alpha}\rightarrow U_{\alpha 1}$, $b_{\alpha}\rightarrow U_{\alpha 2}$, 
$c_{\alpha}\rightarrow U_{\alpha 3}$, and the vacuum oscillation result is reproduced, 
up to the matter induced phase $\delta^L$. 

Second, $P_{ee}$ can be written as 
\begin{equation}
P_{ee}=P_2^H\cos^2\xi(P_{ee}^{2\nu})+P_3^H\sin^2\xi,
\label{pee3} 
\end{equation}
where $P_{ee}^{2\nu}$ is the two-neutrino result obtained in the previous section 
(see Eq.~(\ref{pee})) in the limit $\cos 2\vartheta_M\rightarrow -1$. It is easy to 
check that
Eq.~(\ref{pee}) would be exactly reproduced (with $\vartheta_M$ 
replaced by $\omega_M$, 
of course) if the $\cos2\omega_M= -1$ approximation were dropped. 

For solar neutrino energies (100~keV$\lesssim E_{\nu}\lesssim 10$~MeV), 
$\xi_M\rightarrow\xi$, $P_c^H\rightarrow 0$ and therefore
$P_2^H~(P_3^H)\rightarrow\cos^2\xi~(\sin^2\xi)$, reproducing correctly the result of
the survival probability of electron-type solar neutrinos in a three-flavour 
oscillation scenario (see \cite{general_review} and references therein). 
In this scenario there is no ``$H-L$'' resonance inside the Sun, because
$\frac{|\Delta m^2_{31}|}{2E_{\nu}}\gg A$ for solar neutrino energies. 

On the other hand, in the
case $P_c^H\rightarrow 0$ and $\cos 2\xi_M\rightarrow -1$, $P_3^H\rightarrow 1$ and
electron-type neutrinos exit the Sun as a pure $\nu_3$ mass eigenstate, and do not 
undergo vacuum oscillations even if $\Delta m^2_{21}$ is very small. In contrast,
$\nu_{\mu}$ and $\nu_{\tau}$ always undergo vacuum oscillations if $\Delta m^2_{21}$
is small enough. The reason for this is simple. The generic feature of matter 
effects is to ``push'' $\nu_e$ into the heavy mass eigenstate, while $\nu_{\mu}$ and 
$\nu_{\tau}$
are ``pushed'' into the light mass eigenstates. This situation is changed by 
nonadiabatic effects, as argued above.   

Finally, it is important to note that all equations obtained are also valid in the 
case of inverted hierarchies ($m_3^2<m_{1,2}^2$ or $m_{2}^2<m_1^2$). This has been 
discussed in detail in the two-neutrino oscillation case \cite{earth_matter}, 
and is also applicable here. It is worthwhile to point out that, in the approximation 
$\Delta m^2_{31} \simeq\Delta m^2_{32}$ the transformation $\Delta m^2_{21}
\rightarrow-\Delta m^2_{21}$ can be reproduced by transforming 
$\omega\rightarrow\pi/2-\omega$, $\theta\rightarrow
\pi-\theta$ and redefining the sign of $\nu_{\tau}$. Therefore, one is in principle
allowed to fix the sign of $\Delta m^2_{21}$ as long as $\theta$ is allowed to vary
between $0$ and $\pi$. 

In the case of inverted hierarchies (especially when $\Delta m^2_{31}<0$) one
expects to see no ``level crossing'' (indeed, matter effects tend to increase the
distance between the ``energy'' levels in this case), but matter effects are 
still present, because the initial state in the Sun's core can be nontrivial (
{\it i.e.},\/ $\vartheta_M\neq\vartheta$). Note that $\nu_e$ is still ``pushed''
towards $\nu_H$, even in the case of inverted hierarchies, and the expressions
for the matter mixing angles Eq.~(\ref{cos2tm}), and the initial
states inside the Sun Eq.~(\ref{inistate3}) are still valid. The consequence
of no ``level crossing'' is that the adiabatic limit does not connect, for
example, $\nu_H\rightarrow\nu_3$ but $\nu_H\rightarrow\nu_2$ (or $\nu_1$, depending
on the sign of $\Delta m^2_{21}$). This information is in fact contained in the 
equations above. The crucial feature is that, for example, 
when $\Delta m^2_{31}<0$, $P_c^H\rightarrow 1$ in the ``adiabatic limit,'' and the
matrix $V_{3\times 3}$ correctly ``connects'' $\nu_H\rightarrow\nu_2$ (or $\nu_1$)! 
Another curious feature is that, in the limit
$|\Delta m^2_{31}|/2E_{\nu}\gg A$, $\cos 2\xi_M\rightarrow-\cos2\xi$, 
$P_c^H\rightarrow 1$ and Eq.~(\ref{pee3}) correctly reproduces the survival
probability of electron-type solar neutrinos in the three-flavour oscillation case. 
Note that on this case the sign of $\Delta m^2_{31}$ does not play any role, 
as expected. On the other hand, it is still
true that $P_c^{(H,L)}\rightarrow\cos^2(\xi,\omega)$ in the extreme nonadiabatic 
limit,
and vacuum oscillation results are reproduced, as expected. Again, in this limit, 
one is not sensitive to the sign of $\Delta m^2_{31}$, as expected.  

\subsection{The Case of Antineutrinos}

As in the two-neutrino case, the difference between neutrinos and antineutrinos is
that the equivalent of Eq.~(\ref{eq_3nus}) for antineutrinos can be obtained by
changing $A(x)\rightarrow -A(x)$ and $U_{\alpha i}\leftrightarrow U_{\alpha i}^*$.  
Unlike the two-flavour case, however, there is no set of variable transformations 
that allows one to exactly relate the differential equation for the neutrino and
antineutrino systems. One should, however, 
note that if the signs of both $\Delta m^2$ are
changed and $U_{\alpha i}\leftrightarrow U_{\alpha i}^*$, 
the neutrino equation turns into the antineutrino equation, up to an overall sign. 
This means, for example, that the instantaneous eigenvalues of the antineutrino
Hamiltonian can be read from the eigenvalues of the neutrino Hamiltonian with 
$\Delta m^2_{ij}\leftrightarrow-\Delta m^2_{ij}$, 
$U_{\alpha i}\leftrightarrow U_{\alpha i}^*$ plus an overall sign.

When it comes to computing $P_{\bar{\alpha}\bar{\beta}}$ this global sign
difference is not relevant, and therefore
$P_{\bar{\alpha}\bar{\beta}}(\Delta m^2_{ij},U_{\alpha i})=
P_{\alpha\beta}(-\Delta m^2_{ij},U_{\alpha i}^*)$. 

\setcounter{equation}{0}
\setcounter{footnote}{0}
\section{Results and Discussions}

This section contains the compilation and discussion of a number of results
concerning the oscillation of GeV neutrinos of all species produced in the Sun's
core. The goal here is to explore the multidimensional parameter space spanned by 
$\Delta m^2_{21}$, $\Delta m^2_{31}$, $\sin^2\omega$, $\sin^2\theta$, and 
 $\sin^2\xi$ (and $E_{\nu}$).

It will be assumed throughout that the electron number density profile of the
Sun is exponential, so that Eq.~(\ref{pc}) can be used. As mentioned before,
the numerical accuracy of this approximation is quite good, and certainly good
enough for the purposes of this paper. Therefore, both $P_c^H$ and $P_c^L$ which
appear in Eq.~(\ref{pall3}) will be given by Eqs.~(\ref{pc}, \ref{gamma}), with 
$\vartheta\rightarrow\xi$, $\Delta m^2\rightarrow\Delta m^2_{31}$ in the
former, and $\vartheta\rightarrow\omega$, $\Delta m^2\rightarrow\Delta m^2_{21}$ 
in the latter.

When computing $P_{\alpha\beta}$, an averaging over ``seasons'' is performed, 
which ``washes out'' the effect of very small oscillation wavelengths. 
Furthermore, integration over neutrino energy distributions is performed. Finally,
all $P_{\alpha\beta}$ to be computed should be understood as the value of 
$P_{\alpha\beta}$ in the Earth's surface, {\it i.e.,}\/ Earth matter effects are
not included. This is done in order to make the Sun matter effects in the 
evaluation of $P_{\alpha\beta}$ more clear. It should be stressed that Earth matter
effects may play a significant role for particular regions of the parameter space, but
the discussion of such effects will be left for another opportunity.

Because the parameter space to be explored is multidimensional, it is necessary
to make two-dimensional projections of it, such that ``illustrative'' points are
required. The following points in the parameter space are chosen, all inspired
by the current experimental situation:

\begin{list}{$\bullet$}{}
\item{ATM: $\Delta m^2_{31}=3\times 10^{-3}$~eV$^2$, $\sin^2\theta=0.5$, and
$\sin^2\xi=0.01$,}
\item{LMA: $\Delta m^2_{21}=2\times 10^{-5}$~eV$^2$, $\sin^2\omega=0.2$,}
\item{SMA: $\Delta m^2_{21}=6\times 10^{-6}$~eV$^2$, $\sin^2\omega=0.001$,}
\item{LOW: $\Delta m^2_{21}=1\times 10^{-7}$~eV$^2$, $\sin^2\omega=0.4$,}
\item{VAC: $\Delta m^2_{21}=1\times 10^{-10}$~eV$^2$, $\sin^2\omega=0.55$.}
\end{list}
ATM corresponds to the best fit point of the solution to the atmospheric neutrino
puzzle \cite{atmos_analysis}, 
and a value of $\sin^2\xi=0.01$ which is consistent with all the experimental
bounds. Note that some ``subset'' of ATM will always be assumed (for example, 
$\sin^2\theta$ is fixed while exploring the ($\Delta m^2_{31}\times\sin^2\xi$)-
plane). For each analysis, it will be clear what are the ``variables'' 
and what quantities
are held fixed at their ``preferred point'' values. 
All other points refer to sample points in the regions which best solve
the solar neutrino puzzle \cite{bksreview,rate_analysis,dark_side}, 
and the notation should be obvious. 
Initially a flat neutrino energy distribution with $E_{\nu}^{min}=1$~GeV
and $E_{\nu}^{max}=5$~GeV is considered (for concreteness), 
and the case of higher average energies is briefly discussed later.    

\subsection{The Case of Vacuum Oscillations}

If  neutrinos were produced and propagated exclusively in vacuum, the oscillation
probabilities would be given by Eq.~(\ref{p3vac}). This would be the case of 
neutrinos produced in the Sun's core if either the electron number density were
much smaller than its real value or if very low energy neutrinos were being considered.
Nonetheless, it is still useful to digress some on the ``would be'' vacuum oscillation
probabilities in order to understand better the matter effects.
 
\begin{figure} [t]
\centerline{
  \psfig{file=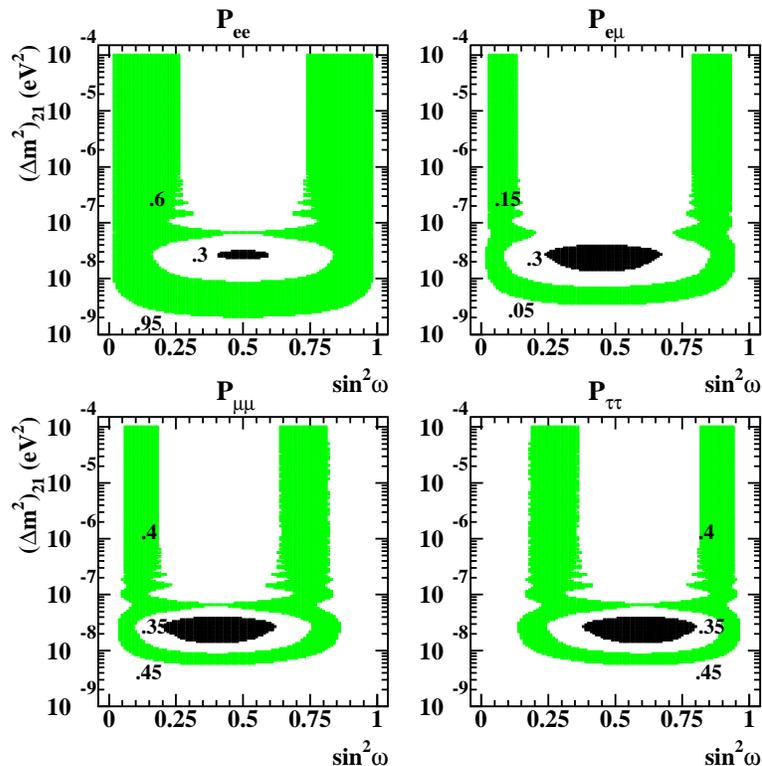,width=0.75\textwidth}
}
\caption{Constant $P_{\alpha\beta}$ contours in the 
($\Delta m^2_{21}\times\sin^2\omega$)-plane, at ATM (see text), in the case of
pure vacuum oscillations.}
\label{dm21_vacuum}
\end{figure}

In the case of pure vacuum oscillations, it is trivial to check that 
$P_{\alpha\beta}=P_{\beta\alpha}$ (remember that the MNS matrix phase 
$\phi$ has been set to zero), and therefore all $P_{\alpha\beta}$ can be parametrised
by three quantities, namely $P_{\alpha\alpha}$, $\alpha=e,\mu,\tau$. It is easy to 
show that
\begin{equation}
P_{\alpha\beta}=P_{\beta\alpha}\Leftrightarrow 
P_{e\mu}=\frac{1}{2}(1+P_{\tau\tau}-P_{\mu\mu}-P_{ee}).
\label{pab=pba}
\end{equation}
From Eq.~(\ref{p3vac})
\begin{eqnarray}
\label{p3vac_aa}
P_{\alpha\alpha}&=&U_{\alpha1}^4+U_{\alpha2}^4+U_{\alpha3}^4+
2U_{\alpha1}^2U_{\alpha2}^2
\cos\left(\frac{\Delta m^2_{21}x}{2E_{\nu}}\right)\nonumber \\ 
&&{\rm or} \\
P_{\alpha\alpha}&=&(1-U_{\alpha3}^2)^2+U_{\alpha3}^4-4U_{\alpha1}^2U_{\alpha2}^2
\sin^2\left(\frac{\Delta m^2_{21}x}{4E_{\nu}}\right). \nonumber 
\end{eqnarray} 
Note that there is no dependency on $\Delta m^2_{31}$. Particularly simple limits can
be reached when $L_{\rm osc}^{21}$ is either very small or very large compared with the
Earth-Sun distance. In both limits $P_{\alpha\alpha}$ is independent of 
$\Delta m^2_{21}$ and, in the latter case, $P_{\alpha\alpha}$ depends only on 
$U_{\alpha3}^2$. Fig.~\ref{dm21_vacuum} depicts constant $P_{\alpha\beta}$ contours
in the  ($\Delta m^2_{21}\times\sin^2\omega$)-plane, at ATM. 
Remember that, here, $P_{e\mu}$ is not an independent quantity
but is a linear combination of all $P_{\alpha\alpha}$.
Note that $P_{ee}$ is symmetric for $\omega\rightarrow\pi/2-\omega$, and that
$P_{\mu\mu}\leftrightarrow P_{\tau\tau}$ when $\omega\rightarrow\pi/2-\omega$. 
The latter property is a consequence of $\theta=\pi/4$. Also, in the case of $P_{ee}$,
the $L_{\rm osc}^{21}\rightarrow\infty$ coincides with the $\omega\rightarrow 0, \pi/2$
limit for any $L_{\rm osc}^{21}$ (this is because either $U_{e1}$ or $U_{e2}$ go to 
zero). This is not true of  $P_{\mu\mu}$ or $P_{\tau\tau}$ unless $\sin^2\xi=0$.
Another important consequence of $L_{\rm osc}^{21}\gg 1$~a.u. is that $T$-violating
effects are absent, even if $\phi$ is nonzero. This can be seen by looking at the
second expression in Eq.~(\ref{p3vac_aa}), which is a function only of 
$|U_{\alpha 3}|^2$ in the limit $L_{\rm osc}^{21}\rightarrow\infty$. 
  
Finally, one should note that oscillatory effects are maximal for 
$\Delta m^2_{21}\simeq 2\times 10^{-8}$~eV$^2$. In this region $
\cos\left(\frac{\Delta m^2_{21}x}{2E_{\nu}}\right)\simeq -1$, and the largest 
suppression to all $P_{\alpha\alpha}$ is obtained when $U_{\alpha1}^2U_{\alpha2}^2$
is maximum. For example, $P_{ee}$ is smallest when $\omega=\pi/4$, 
since $U_{e1}^2U_{e2}^2\propto\sin^2 2\omega$. 
There are no ``localised'' maxima for $P_{\alpha\alpha}$ because 
$U_{\alpha1}^2U_{\alpha2}^2$ is positive definite. 

\subsection{``Normal'' Neutrino Hierarchy}

When matter effects are ``turned on,'' the situation can be dramatically different.
This is especially true in the case of normal neutrino mass hierarchies 
($m_1^2<m_2^2<m_3^2$), which will be discussed first. 
 
\begin{figure} [t]
\centerline{
  \psfig{file=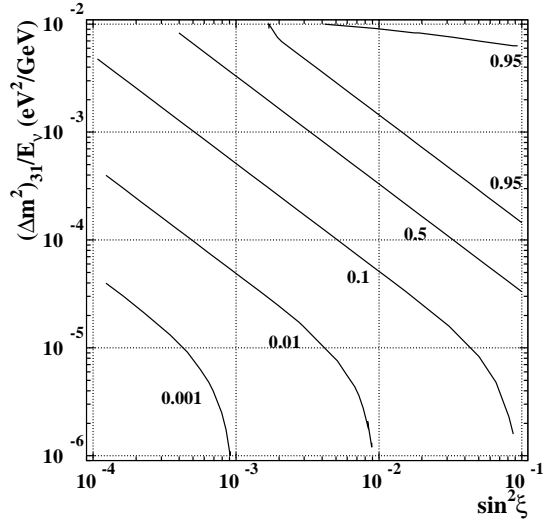,width=0.5\textwidth}
}
\caption{Contours of constant $P_3^H$ (see text) in the 
($\Delta m^2_{31}/E_{\nu}\times\sin^2\xi$)-plane.}
\label{p3h}
\end{figure}

The first effect one should observe is that, even though $L_{\rm osc}^{31}\ll 1$~a.u.,
$P_{\alpha\beta}$ depend rather nontrivially on $\Delta m^2_{31}$. This dependency
comes from the terms $P_3^H$ and $P_2^H=1-P_3^H$ in Eq.~(\ref{pall3}). Remember that
$P_3^H$ is interpreted as the probability that a $\nu_e$ produced in the Sun's core
exits the Sun as a $\nu_3$ mass eigenstate. When matter effects are negligible 
(such as in the limit of small neutrino energies) $P_3^H\rightarrow\sin^2\xi$, its
``vacuum limit.'' Fig.~\ref{p3h} depicts constant $P_3^H$ contours in the 
($\Delta m^2_{31}/E_{\nu}\times\sin^2\xi$)-plane.  

Note that, for $\Delta m^2_{31}/E_{\nu}\sim 10^{-2}$~eV$^2$/GeV, 
$P_3^H\rightarrow 1$, even
for small values of $\sin^2\xi$. In this region, $\nu_e$'s produced in the Sun's
core exit the Sun as pure $\nu_3$'s. Therefore, $P_{e\alpha}\simeq U_{\alpha3}^2$. 
Because of unitarity in the propagation, $\nu_{\mu}$'s
and $\nu_{\tau}$'s exit the Sun as linear combinations of the light mass eigenstates,
and may not only undergo vacuum oscillations but are also susceptible to further
matter effects (dictated by the ``$M-L$'' system, as described in Sec.~3). For future
reference, at ATM, $P_3^H\simeq0.87$ when averaged over the energy range mentioned 
in the beginning of this section.  

As $\Delta m^2_{31}/E_{\nu}$ decreases (as is the case for higher energy neutrinos)
the nonadiabaticity of the ``$H-M$'' system starts to become relevant, and 
$P_3^H\rightarrow\sin^2\xi$, as argued in Sec.~3.2. A hint of this behaviour can 
already be seen in Fig.~\ref{p3h}, for small values of $\Delta m^2_{31}/E_{\nu}$.

\begin{figure} [t]
\centerline{
  \psfig{file=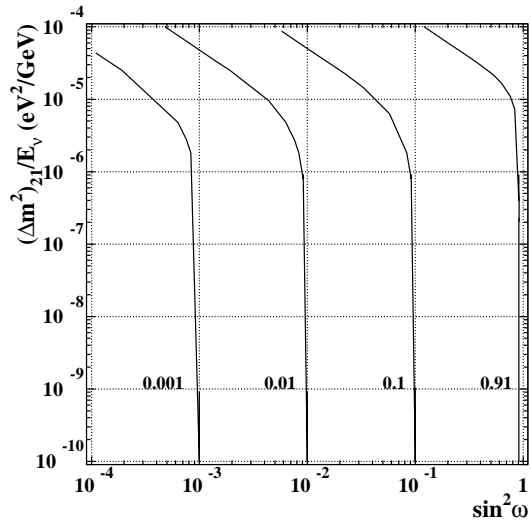,width=0.5\textwidth}
}
\caption{Contours of constant $1-P_c^L$ (see text) in the 
($\Delta m^2_{21}/E_{\nu}\times\sin^2\omega$)-plane.}
\label{1-pcl}
\end{figure}

The information due to the ``$M-L$'' matter effect is encoded in $P_c^L$, present
in Eq.~(\ref{pall3}). Fig.~\ref{1-pcl} depicts contours of constant $1-P_c^L$ in 
the ($\Delta m^2_{21}/E_{\nu}\times\sin^2\omega$)-plane. One should note that 
$1-P_c^L$ reaches its extreme
nonadiabatic limit, $\sin^2\omega$, when $\Delta m^2_{21}/E_{\nu}\lesssim 
10^{-7}$~eV$^2$/GeV. For $\Delta m^2_{21}/E_{\nu}\gtrsim 
10^{-7}$~eV$^2$/GeV, matter effects increase the value of $1-P_c^L$. 

One can use the intuition from the two-flavour solution to the solar neutrino puzzle
to better appreciate the results presented here. 
In the case of the solutions to the solar neutrino puzzle, 
the energies of interest range from 100~keV to 10~MeV, and large matter effects happen 
around $\Delta m^2\sim 10^{-5}$~eV$^2$. Furthermore, at 
$\Delta m^2\sim 10^{-10}$~eV$^2$ one encounters the ``just-so'' solution, which is 
characterised by very long wave-length vacuum oscillations. Rescaling to $O$(GeV) 
energies, the equivalent of the ``just-so'' solution happens for 
$\Delta m^2_{21}\sim (10^{-8}-10^{-7})$~eV$^2$, while large matter effects would 
be present at $\Delta m^2\sim (10^{-3}-10^{-2})$~eV$^2$. Indeed, one observes
large matter effects for $\Delta m^2_{31}\sim (10^{-3}-10^{-2})$~eV$^2$. 
$\Delta m^2_{21}\sim (10^{-5}-10^{-6})$~eV$^2$ 
corresponds to the region between the LOW and VAC 
solutions, where matter effects distort $P_{\alpha\beta}$ from its pure vacuum value, 
but no dramatic suppression or enhancement takes place. Incidently, this behaviour 
has physical consequences in the solution to the solar neutrino problem, as was 
first pointed out in \cite{alex}. 

\begin{figure} [pt]
\centerline{
  \psfig{file=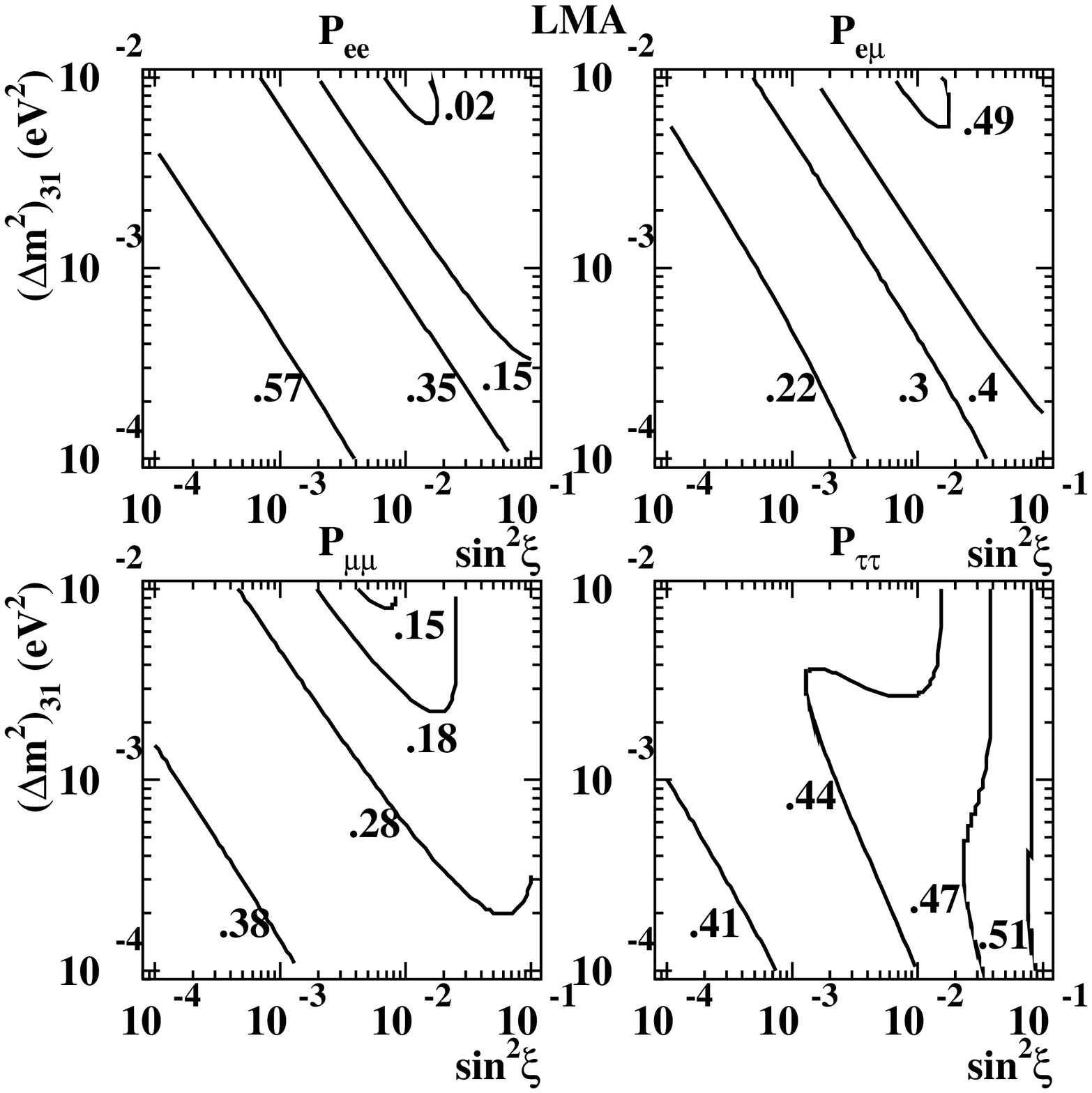,width=0.6\textwidth}}
\centerline{
  \psfig{file=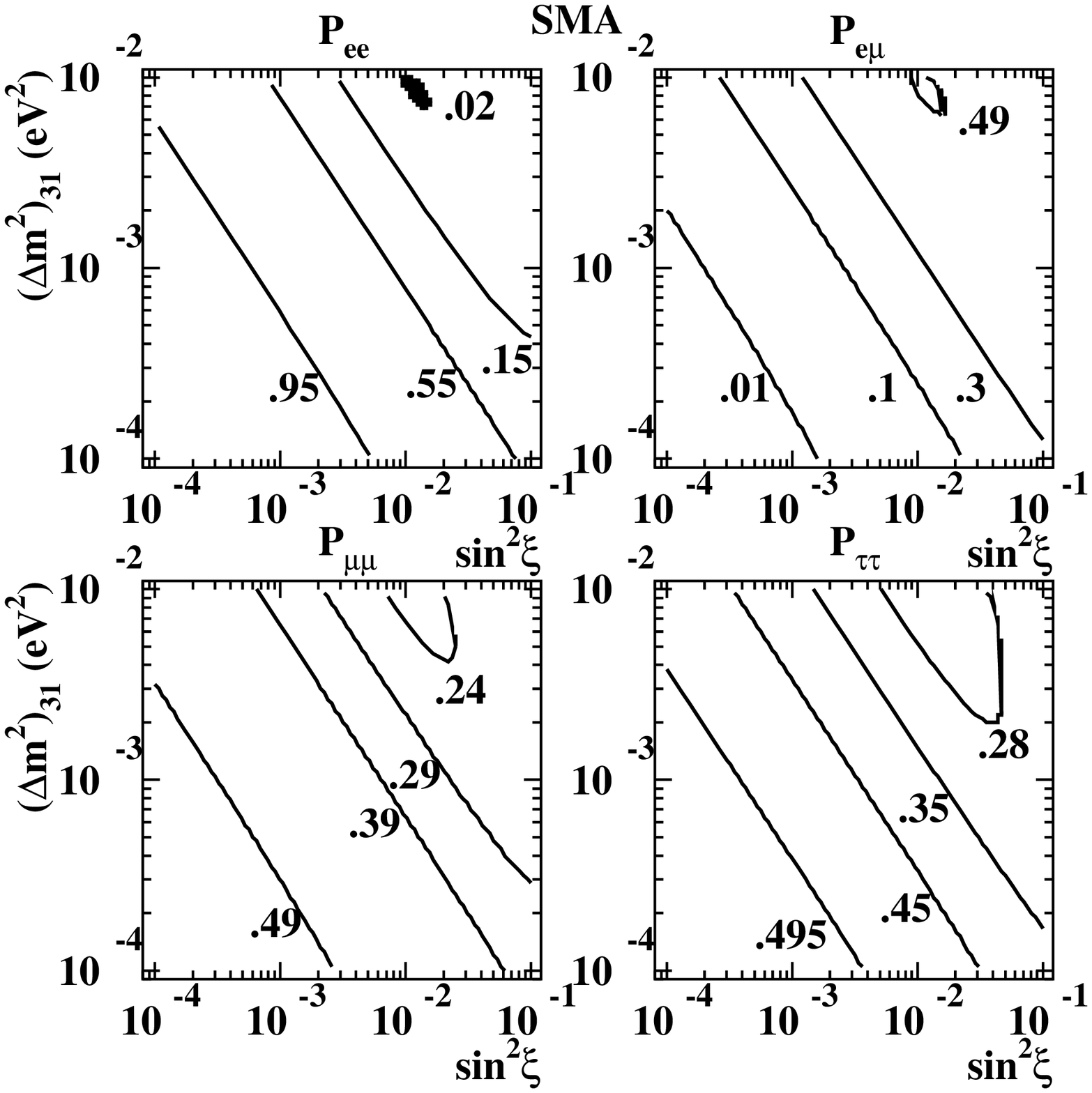,width=0.6\textwidth}}
\caption{Constant $P_{\alpha\beta}$ contours in the 
($\Delta m^2_{31}\times\sin^2\xi$)-plane, at LMA (top) and SMA (bottom) (see text).}
\label{dm31_ssxi_lma}
\end{figure}

\begin{figure} [pt]
\centerline{
  \psfig{file=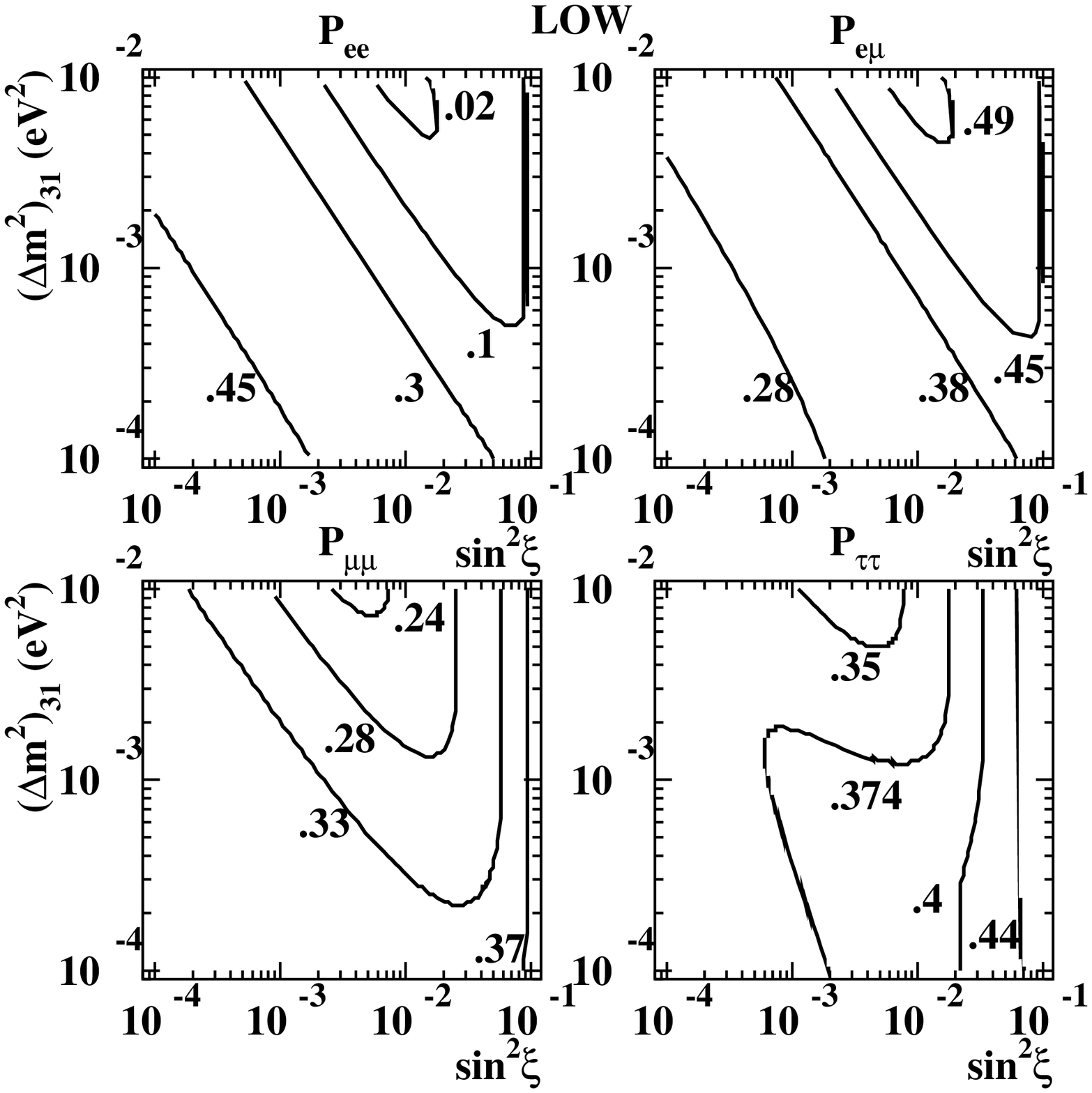,width=0.6\textwidth}}
\centerline{
  \psfig{file=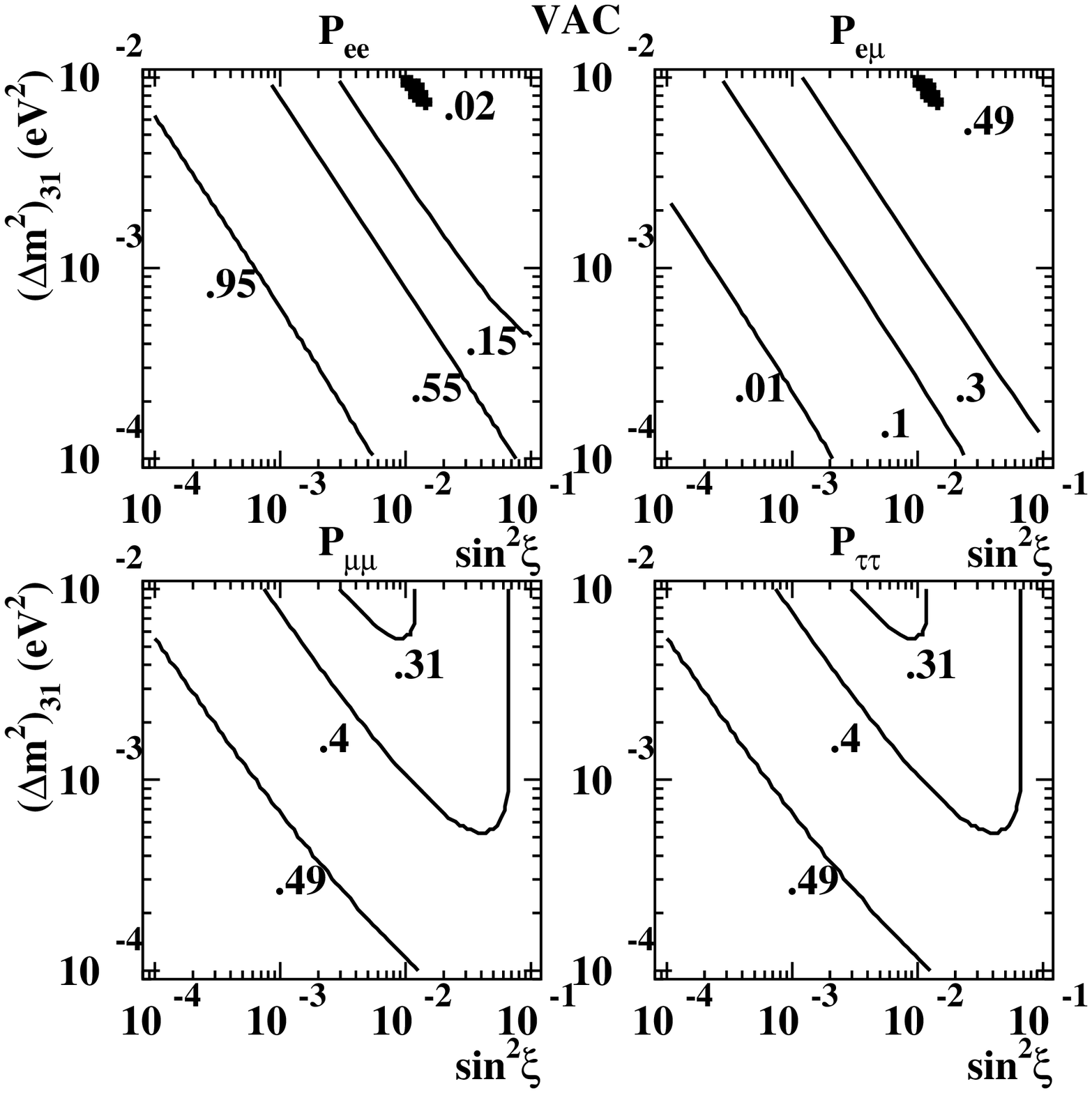,width=0.6\textwidth}}
\caption{Constant $P_{\alpha\beta}$ contours in the 
($\Delta m^2_{31}\times\sin^2\xi$)-plane, at LOW (top) and VAC (bottom) (see text).}
\label{dm31_ssxi_low}
\end{figure}

Figs.~\ref{dm31_ssxi_lma} and \ref{dm31_ssxi_low} depict contours of constant
$P_{\alpha\alpha}$ and $P_{e\mu}$ in the ($\Delta m^2_{31}\times\sin^2\xi$)-plane.
As expected, in the region where $P_3^H\sim 1$, $P_{ee}$ and $P_{e\mu}$ do not
depend on $\Delta m^2_{21}$ or $\sin^2\omega$, namely 
$P_{ee}\sim\sin^2\xi$ and $P_{e\mu}\sim 0.5\cos^2\xi$. Remember that the results 
depicted in Figs.~\ref{dm31_ssxi_lma} and \ref{dm31_ssxi_low}
(and all other plots from here on)
are for an energy band from 1 to 5~GeV. On the other hand, 
$P_{\mu\mu}$ and $P_{\tau\tau}$ do depend on the point (LMA, SMA, etc), even for
$P_3^H\sim 1$, as foreseen. This dependence will be discussed in what follows.
    
In the limit $P_3^H=1$, $\sin^2\theta=0.5$ 
\begin{eqnarray}
c_{\mu}^2=&c_{\tau}^2=&0, \\
a_{\mu}^2=&b_{\tau}^2=&0.5(1+2\sqrt{P_c^L(1-P_c^L))}, \\
b_{\mu}^2=&a_{\tau}^2=&0.5(1-2\sqrt{P_c^L(1-P_c^L))}, \\
\label{amubmu}
a_{\mu}b_{\mu}=&-a_{\tau}b_{\tau}=&0.5(1-2P_c^L),
\end{eqnarray}
and 
\begin{eqnarray}
\label{pmmtt}
P_{(\mu\mu,\tau\tau)}&=&\frac{1}{2}(1-U_{(\mu,\tau)3}^2)\pm\sqrt{P_c^L(1-P_c^L)}
(U_{(\mu,\tau)1}^2-U_{(\mu,\tau)2}^2) \\ &\pm& (1-2P_c^L)U_{(\mu,\tau)1}U_{(\mu,\tau)2}
\cos\left(\frac{\Delta m^2_{21}x}{2E_{\nu}}\right). \nonumber
\end{eqnarray}
At both LMA and SMA, the oscillatory term averages out to zero, while at VAC
$\cos\left(\frac{\Delta m^2_{21}x}{2E_{\nu}}\right)=1$. It is
only at LOW that the oscillatory term is nontrivial, as was mentioned in the analogy
between the situation at hand and the solutions to the solar neutrino puzzle.
 
Furthermore, at SMA, $1-P_c^L$ is tiny (see Fig.~\ref{1-pcl}), 
so it is fair to approximate $P_{\mu\mu}\simeq P_{\tau\tau}\simeq 0.5(1-0.5\times0.99)
\simeq 0.25$, in agreement with Fig.~\ref{dm31_ssxi_lma}(bottom). 
At LMA, it is fair to
approximate $\sin^2\xi=0$. In this limit, 
$U_{\mu1}^2-U_{\mu2}^2\simeq -0.5\cos 2\omega+\sin 2\omega\sin\xi$, while
$U_{\tau1}^2-U_{\tau2}^2\simeq -0.5\cos 2\omega-\sin 2\omega\sin\xi$. 
Therefore, because
$\cos 2\omega=0.6>0$, $P_{\mu\mu}$ is significantly less than $P_{\tau\tau}$, since
$\sqrt{P_c^L(1-P_c^L)}$ is nonnegligible. 
Roughly, $P_{\mu\mu}\simeq 0.15$ and $P_{\mu\mu}\simeq 0.4$, using the 
approximations above. Again, there is agreement with Fig.~\ref{dm31_ssxi_lma}(top).

In order to understand the behaviour at LOW and VAC, one should take advantage of
the fact that $1-P_c^L\rightarrow\sin^2\omega$. In this case, it proves more 
advantageous to use the second form of Eq.~(\ref{pall3}) in order to express all
$P_{\alpha\beta}$
\begin{eqnarray}
\label{1-pcl=sso}
P_{ee}&=&P_2^H\cos^2\xi+P_3^H\sin^2\xi-{\rm (Osc)}_{ee}, \nonumber \\
P_{e\mu}&=&\left(P_2^H\sin^2\xi+P_3^H\cos^2\xi\right)\sin^2\theta-
{\rm (Osc)}_{e\mu}, \\ 
P_{\mu\mu}&=&\left(\cos^2\theta+\sqrt{P_3^H}\sin\xi\sin^2\theta\right)^2
+P_2^H\cos^2\xi\sin^4\theta-{\rm (Osc)}_{\mu\mu}, \nonumber \\
P_{\tau\tau}&=&\left(\sin^2\theta+\sqrt{P_3^H}\sin\xi\cos^2\theta\right)^2
+P_2^H\cos^2\xi\cos^4\theta-{\rm (Osc)}_{\tau\tau}, \nonumber
\end{eqnarray}
where
\begin{equation}
{\rm (Osc)}_{\alpha\beta}=4a_{\alpha}b_{\alpha}U_{\beta1}U_{\beta2}
\sin^2\left(\frac{\Delta m^2_{12}x}{4E_{\nu}}\right)
\end{equation}
are the oscillatory terms. When $L_{\rm osc}^{21}\gg 1$~a.u., the oscillatory
terms are zero, and $P_{\alpha\beta}$ are particularly simple. Note that on this limit
many simplifications happen: $P_{\alpha\beta}$ is independent of $\omega$ and
$\Delta m^2_{21}$, and $P_{\mu\mu}=P_{\tau\tau}$ if $\sin^2\theta=\cos^2\theta$, as
can be observed in Fig.~\ref{dm31_ssxi_low}(bottom). 
A very important fact is that, when the
oscillatory terms are neglected, $2P_{e\mu}=1+P_{\tau\tau}-P_{\mu\mu}-P_{ee}$, as
one may easily verify directly. As argued before, when this condition is satisfied,
$P_{\alpha\beta}=P_{\beta\alpha}$. This is not the case in the presence of nonnegligible
oscillation effects or when $P_c\neq\cos^2\omega$. Both statements are trivial to
verify directly. For example, 
\begin{equation}
\label{abmumu}
4a_e b_e U_{\mu1} U_{\mu2}=P_2^H\sin 2\omega\left[\sin 2\omega\left(\sin^2\xi
\sin^2\theta-\cos^2\theta\right)-\sin\xi\sin 2\theta\cos 2\omega\right]
\end{equation}
while
\begin{equation}
\label{abee}
4a_{\mu} b_{\mu} U_{e1} U_{e2}=\cos^2\xi\sin 2\omega\left[\sin 2\omega
\left(P_3^H\sin^2\theta-\cos^2\theta\right)-
\sqrt{P_3^H}\sin 2\theta\cos 2\omega\right],
\end{equation}
so ${\rm Osc}_{e\mu}\neq{\rm Osc}_{\mu e}$. 

\begin{figure} [pt]
\centerline{
  \psfig{file=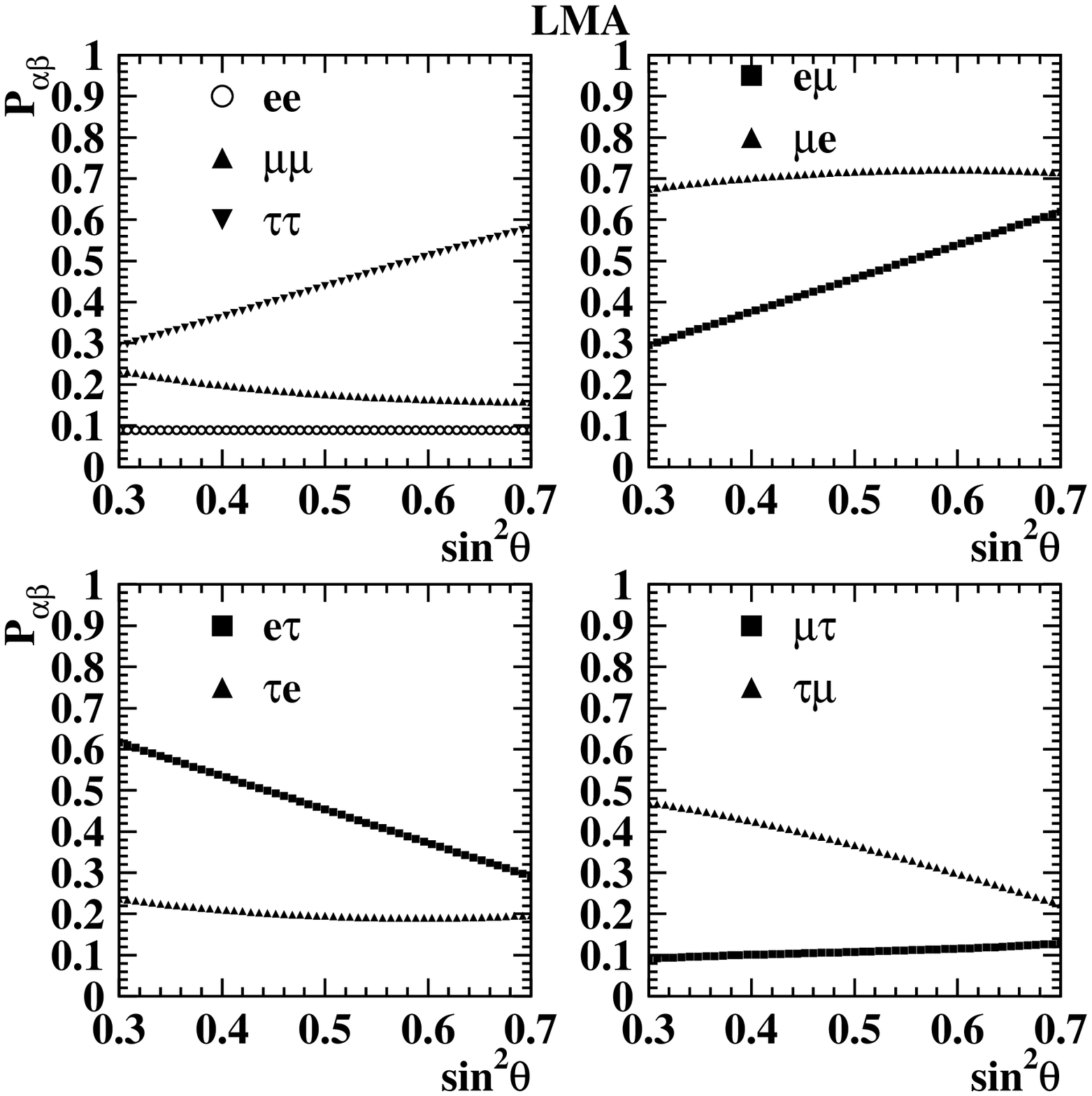,width=0.6\textwidth}}
\centerline{
  \psfig{file=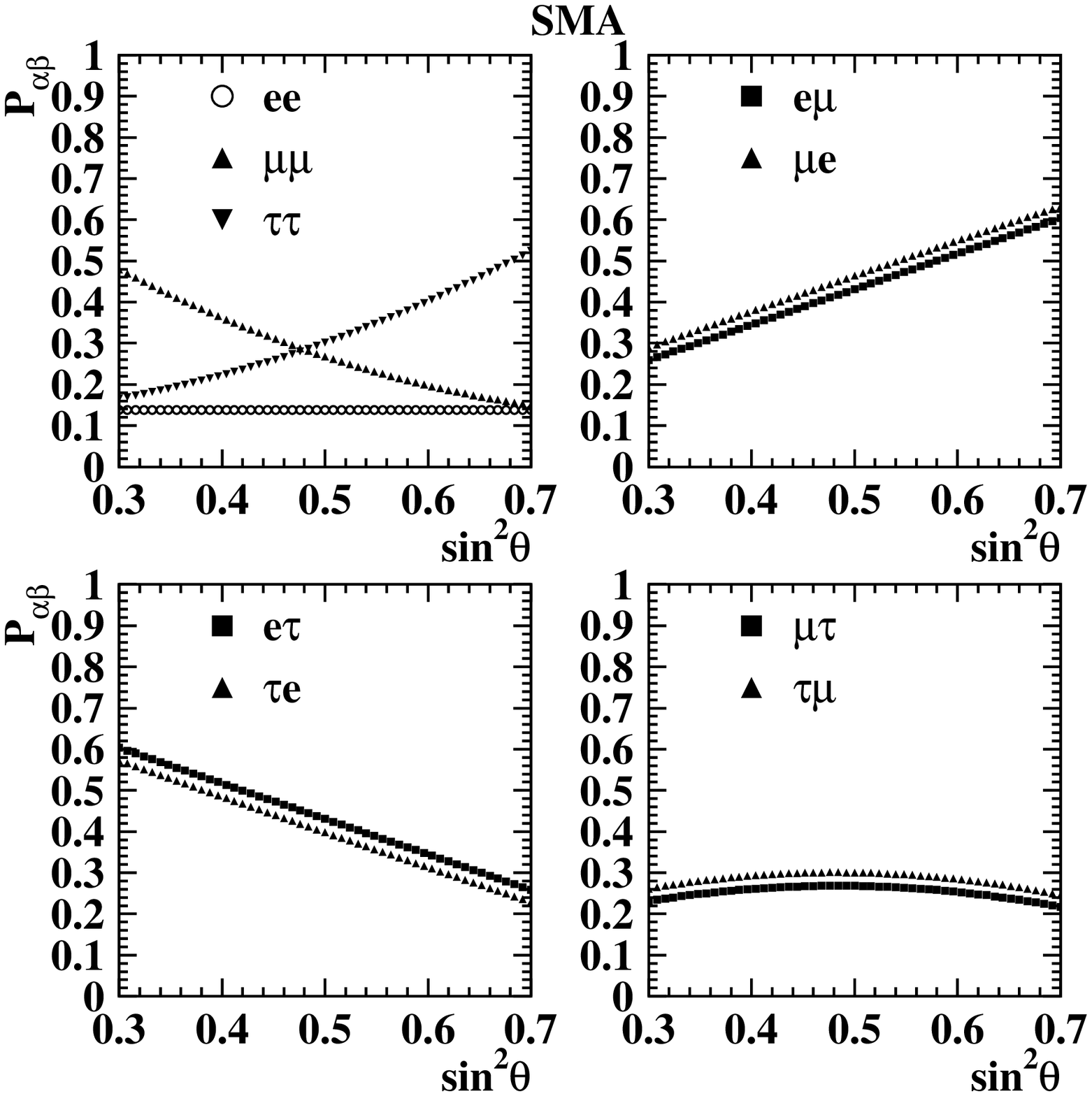,width=0.6\textwidth}}
\caption{$P_{\alpha\beta}$ as a function of $\sin^2\theta$, at LMA (top) and
SMA (bottom) (see text).}
\label{sst_lma}
\end{figure}

\begin{figure} [pt]
\centerline{
  \psfig{file=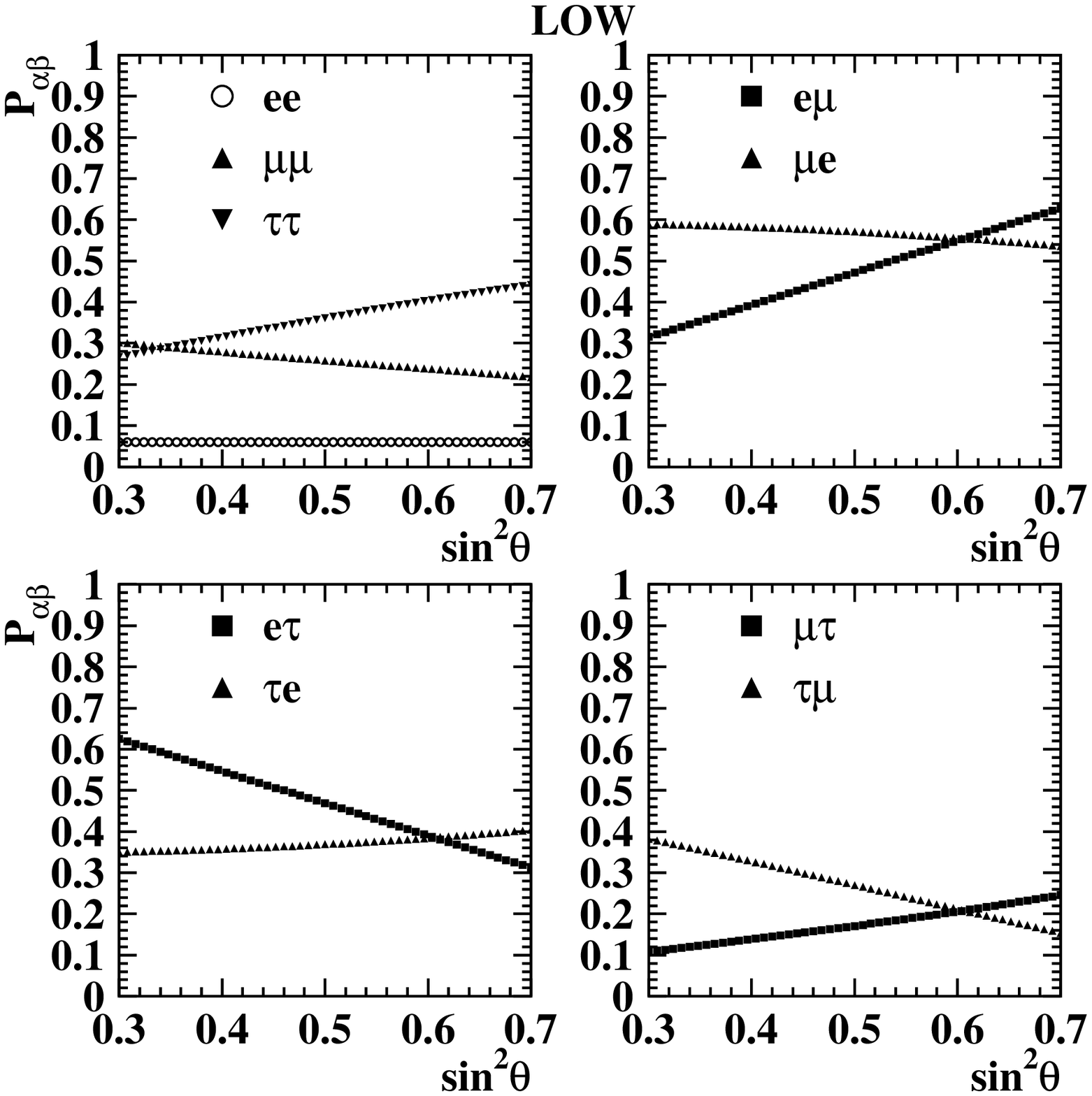,width=0.6\textwidth}}
\centerline{
  \psfig{file=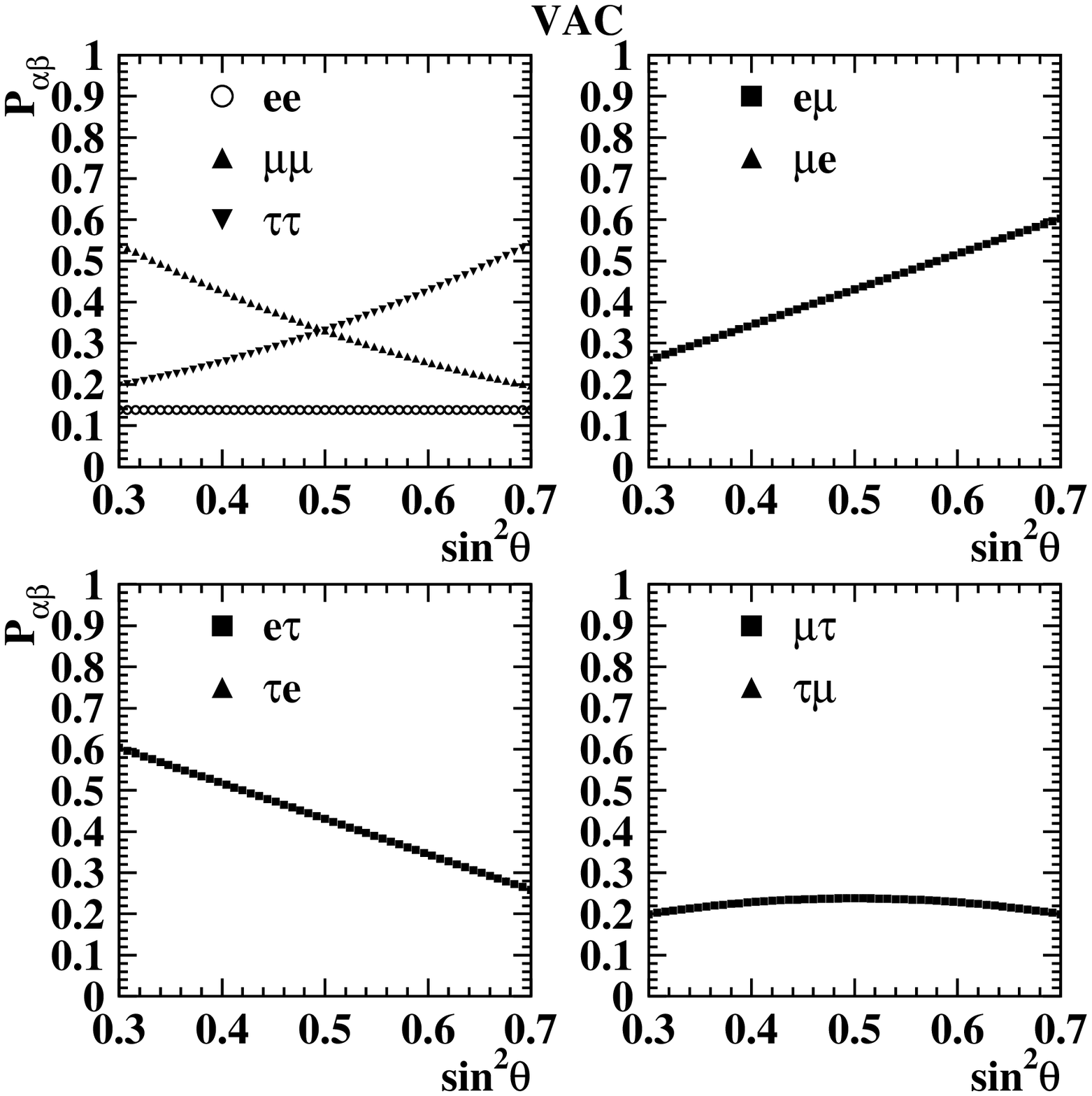,width=0.6\textwidth}}
\caption{$P_{\alpha\beta}$ as a function of $\sin^2\theta$, at LOW (top) and
VAC (bottom) (see text).}
\label{sst_low}
\end{figure}

Figs.~\ref{sst_lma} and \ref{sst_low} depict $P_{\alpha\beta}$ as a function of
$\sin^2\theta$ at LMA and SMA, and at LOW and VAC, respectively. In these figures, all 
$P_{\alpha\beta}$ are plotted, in order to illustrate that $P_{\alpha\beta}\neq
P_{\beta\alpha}$ at LMA, SMA and VAC. Note that at LMA and SMA, the difference comes
from the fact that $P_3^H\neq \sin^2\xi$ {\sl and}\/ $P_c^L\neq \cos^2\omega$. At
LOW, $P_c^L\simeq\cos^2\omega$, but $P_3^H\neq \sin^2\xi$ {\sl and} 
nontrivial oscillatory terms render $P_{\alpha\beta}\neq P_{\beta\alpha}$. 
At VAC, even though $P_3^H\neq \sin^2\xi$, $P_{\alpha\beta}=P_{\beta\alpha}$ because
$P_c^L=\cos^2\omega$ {\sl and} because ``1-2'' oscillations don't have ``time''
to happen. 


From Eqs.~(\ref{1-pcl=sso}) one can roughly understand the dependency of
$P_{\alpha\beta}$ on $\sin^2\theta$. Obviously $P_{ee}$ does not depend on $\theta$
(by the very form of the MNS matrix, Eq.~(\ref{MNSmatrix})), 
while $P_{e\mu}$ ($P_{e\tau}$) depends almost
exclusively on $\sin^2\theta$ ($\cos^2\theta$). 
This is guaranteed by the fact that $P_3^H\gg P_2^H$
even at LMA and LOW, when one expects the interference terms to play a 
significant role. It is also worthwhile to note that, as expected, at VAC and SMA 
the curves are very similar, a behaviour that can be understood from earlier 
discussions.   

\begin{figure} [t]
\centerline{
  \psfig{file=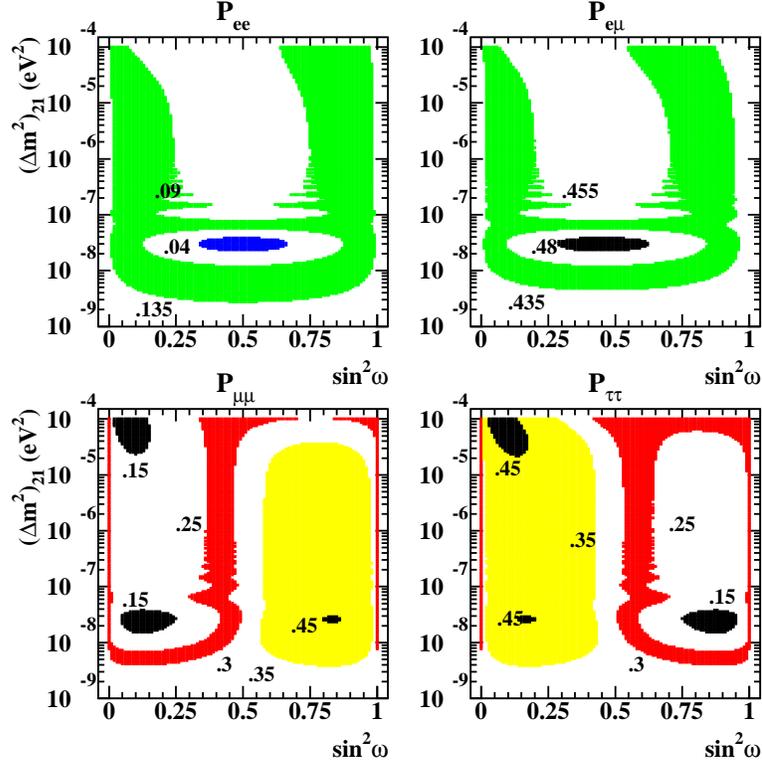,width=0.75\textwidth}
}
\caption{Constant $P_{\alpha\beta}$ contours in the 
($\Delta m^2_{21}\times\sin^2\omega$)-plane, at ATM (see text).}
\label{dm21_std}
\end{figure}

Finally, Fig.~\ref{dm21_std} depicts constant $P_{\alpha\beta}$ contours in 
the ($\Delta m^2_{21}\times\sin^2\omega$)-plane, at ATM. In light of the previous 
discussions, the shapes and forms can be readily understood. First note that
the shapes of the constant $P_{ee}$ and $P_{e\mu}$ regions resemble those of 
the pure vacuum oscillations depicted in Fig.~\ref{dm21_vacuum}, with two important 
differences. First, the constant values of the contours are quite different.
For example, $P_{ee}$ varies from a few percent to less then 15\%, while in the 
case of pure vacuum oscillations, $P_{ee}$ varies from 30\% to 100\%. This can be
roughly understood numerically by noting that 
$P_{(ee,e\mu)}\simeq P_2^H P_{(ee,e\mu)}^{\rm vac}$ (remember that $P_2^H=1-P^H_3\simeq 
0.13$ when averaged over the energy range of interest). Second, at
high $\Delta m^2_{21}$, the regions are distorted. This is due to nontrivial 
matter effects in the ``$M-L$'' system. Note that the contours follow the 
constant $1-P_c^L$ curves depicted in Fig.~\ref{1-pcl}.

The $P_{\mu\mu}$ and $P_{\tau\tau}$ contours are a lot less familiar, and require 
some more discussion. Many features are rather prominent. For example, the plane is
roughly divided into a $\sin^2\omega>0.5$ and $\sin^2\omega<0.5$ structure, and 
large (small) values of $P_{\mu\mu}$ ($P_{\tau\tau}$) are constrained to the 
$\sin^2\omega>0.5$ half, and vice-versa. Also, there is a rough $P_{\mu\mu}(\omega)
\leftrightarrow P_{\tau\tau}(\pi/2-\omega)$ symmetry in the picture, which was 
present in the pure vacuum case (see Fig.~\ref{dm21_vacuum}). This symmetry is absent
for large values of $\Delta m^2_{21}$, similar to what happens in the case of $P_{ee}$,
and is due, as mentioned in the previous paragraph, to the fact that 
$P_c^L$ is significantly different from $\cos^2\omega$ in this region.

The other features are also fairly simple to understand, and are all due to fact that
$P_3^H\gg\sin^2\xi$. It is convenient to start the discussion in the limit when 
$L_{21}^{\rm osc}\gg 1$~a.u. (the very small $\Delta m^2_{21}$ region). As was noted 
before, $P_{\alpha\beta}$ are given by Eq.~(\ref{1-pcl=sso}) where the 
Osc$_{\alpha\beta}$ terms vanish. It is therefore easy to see that 
$P_{\alpha\beta}$ do not depend on $\omega$ or $\Delta m^2_{21}$ (as mentioned before),
and furthermore it is trivial to compute the value of $P_{\alpha\beta}$ given that we
are at ATM and that $P_3^H\simeq 0.87$. The next curious feature is that there is a
``band'' around $\sin^2\omega=1/2$ where $P_{\mu\mu}\simeq 
P_{\mu\mu}(L_{21}^{\rm osc}\rightarrow\infty)$. The same is true
of $P_{\tau\tau}$. This is due to the fact that, around $\sin^2\omega\simeq 1/2$,
$a_{\mu}b_{\mu}$ and $a_{\tau}b_{\tau}$ vanish when $P_3^H$ is large. In the limit
$P_3^H=1$ one can use Eq.(\ref{amubmu}) and note that indeed both $a_{\mu}b_{\mu}$
and $a_{\tau}b_{\tau}$ vanish at $P_c^L=1/2$. However, for values of 
$\Delta m^2_{21}\lesssim 10^{-7}$~eV$^2$ $P_c^L\simeq\cos^2\omega$, which explains the
band around $\omega\simeq \pi/4$. Slight distortions are due to the fact that 
$P_3^H\neq1$, and are easily computed from the exact expressions. 

Again in the limit $P_3^H=1$, $P_c^L=\cos^2\omega$, the coefficient of the
$\cos\left(\frac{\Delta m^2_{21}x}{2E_{\nu}}\right)$ term Eq.~(\ref{pmmtt}) is
\begin{equation}
\pm\frac{1}{2}(1-2\cos^2\omega)\left(-\frac{1}{2}\sin 2\omega\mp 0.1\cos 2\omega\right),
\end{equation}
if $\sin^2\xi$ terms are neglected. The $+,-$ signs are for $P_{\mu\mu}$ while the
$-,+$ signs for $P_{\tau\tau}$. It is trivial to verify numerically (if a little 
tedious) that the $P_{\mu\mu}$ term has a maximum at $\sin^2\omega\simeq 0.1$ and a 
minimum at $\sin^2\omega\simeq 0.8$. For $P_{\tau\tau}$ the maximum (minimum) is 
at $\sin^2\omega\simeq 0.9 (0.2)$. It is important to comment that the minima are 
negative numbers. On the other hand, from Fig.~\ref{dm21_vacuum} 
(as mentioned before) it is easy to see that 
$\cos\left(\frac{\Delta m^2_{21}x}{2E_{\nu}}\right)$ is minimum
for $\Delta m^2_{21}\simeq 2\times 10^{-8}$~eV$^2$ (this is where all 
$P_{\alpha\alpha}$ are maximally suppressed in Fig.~\ref{dm21_vacuum}). Combining both
informations, it is simple to understand the maxima/minima of $P_{\mu\mu}$ and 
$P_{\tau\tau}$ at $\Delta m^2_{21}\simeq 2\times 10^{-8}$~eV$^2$: Minima occur
when the coefficient is maximum ({\it e.g.,}\/ at 
$\sin^2\omega\simeq 0.1$ for $P_{\mu\mu}$) while maxima occur when the coefficient is 
minimum ({\it e.g.,}\/ at $\sin^2\omega\simeq 0.8$ for $P_{\mu\mu}$). A description
of what has happened is the following: The matter effects ``compress'' the constant 
$P_{\mu\mu}$ ($P_{\tau\tau}$) contours from the pure vacuum oscillation case 
(presented in Fig.~\ref{dm21_vacuum}) to the $\sin^2\omega<1/2$ ($>1/2$) half of the 
plane, and a new region ``appears'' on the other half. This other region is 
characterised by negative values to the coefficients of the oscillatory terms, which 
are not attainable in the case of pure vacuum oscillations  
(see Eq.~(\ref{p3vac_aa})). 

At last, the contours in the region where the oscillatory effects average out, 
$P_{\mu\mu}$ and $P_{\tau\tau}$ are also best understood from Eq.~(\ref{pmmtt}) and
the paragraphs which follow it, in
the limit that $\cos\left(\frac{\Delta m^2_{21}x}{2E_{\nu}}\right)\rightarrow 0$. 
It is simple to see, for example, that $P_{\mu\mu}<P_{\tau\tau}$ if $\cos 2\omega>0$
($\sin^2\omega>1/2$), while the situation is reversed if $\cos 2\omega<0$. This is
indeed what one observes in Fig.~\ref{dm21_std}.

\subsection{``Inverted'' Neutrino Hierarchy}

Here I turn to the case of an ``inverted'' neutrino hierarchy, namely 
$\Delta m^2_{31}<0$. Currently, there is no experimental hint as to what the sign 
of $\Delta m^2_{31}$ should be, so there is no reason to believe that the ``normal'' 
hierarchy is to be preferred over the ``inverted'' hierarchy. Indeed, even from a 
theoretical/ model building point of view, there are no strong reasons for or against 
a particular neutrino mass hierarchy \cite{theory_review}.  

The discussion will be restricted to $\Delta m^2_{21}>0$ for two
reasons. First, the $\Delta m^2_{21}<0$ can be approximately read off from the 
$\Delta m^2_{21}>0$ case by changing $\omega\rightarrow\pi/2-\omega$, as mentioned
before. Second, and most important, there is some experimental hints as to what is
the sign of $\Delta m^2_{21}$ \cite{solar_3,dark_side}. 
For example, the SMA solution only exists for one
sign of $\Delta m^2_{21}$, while the LMA and LOW solutions prefer 
one particular sign.
Even in the case of VAC there is the possibility of obtaining information concerning 
the sign of $\Delta m^2_{21}$ from solar neutrino data \cite{alex}. 
Therefore, the notation introduced in the beginning
of this section (ATM, SMA, LMA, LOW, VAC) still applies, and one should simply
remember that here $\Delta m^2_{31}<0$.

As advertised, the largest effect of $\Delta m^2_{31}<0$ is the typical values of
$P_{c}^H$. From Eq.~(\ref{pc}), keeping in mind that here $\gamma$ is negative, 
\begin{equation}
P_c^H=\frac{1-e^{-|\gamma|\cos^2\xi}}{1-e^{-|\gamma|}},
\end{equation}
where $\gamma$ is given by Eq.~(\ref{gamma}) with $\Delta m^2\rightarrow\Delta m^2_{31}$.
Since $|\gamma|\gg 1$ (see Eq.~(\ref{gamma})), $P_c^H=1$ for all values of 
$\Delta m^2_{31}$ and $\sin^2\xi$ of interest. Indeed, this is true for any value of
$\sin^2\xi$ as long as $\Delta m^2_{31}\gg 10^{-6}$~eV$^2$. This is to be contrasted to
the normal hierarchy case, when there is always some value of $\sin^2\xi$ (which is a 
function of $\Delta m^2_{31}$) below which
$P_c$ deviates significantly from its adiabatic limit.

All of the $\Delta m^2_{31}$ dependency of $P_{\alpha\beta}$ is therefore encoded in
$\xi_M$. However, in the case $\Delta m^2_{31}<0$ it is trivial to show that 
$-1<\cos 2\xi_M<-\cos 2\xi$, where the upper bound is reached in the limit 
$|\Delta m^2_{31}/2E_{\nu}|\gg A$ 
(this has been mentioned before. The minus sign takes care
of the ``unorthodox'' $P_c\rightarrow 1$ adiabatic limit). Since one is interested in
$\sin^2\xi<0.1$ ($-\cos 2\xi<-0.8$), 
the range for $\xi_M$ is rather limited, and therefore any 
$\Delta m^2_{31}$ effects are bound to be very small. Larger $\Delta m^2_{31}$ effects 
are expected for larger $\sin^2\xi$.

\begin{figure} [t]
\centerline{
  \psfig{file=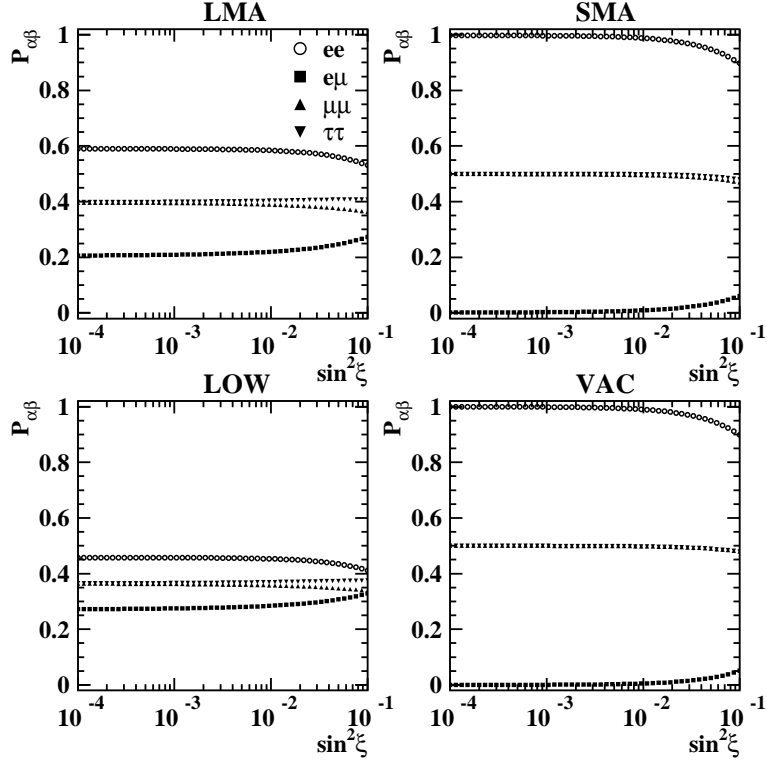,width=0.75\textwidth}
}
\caption{$P_{\alpha\beta}$ as a function of $\sin^2\xi$, at LMA, SMA, LOW, and VAC 
(see text), for $\Delta m^2_{31}<0$.}
\label{ssxi_min}
\end{figure}

In light of this, Fig.~\ref{ssxi_min} depicts $P_{\alpha\beta}$ and $P_{e\mu}$ as
a function of $\sin^2\xi$ at the various points (LMA, SMA, LOW, VAC), for 
$\Delta m^2_{31}=-3\times 10^{-3}$~eV$^2$ and $\sin^2\theta=0.5$.

It is interesting to compare the results presented here with the pure vacuum case. 
In the limit $P_2^H=1$ 
\begin{equation}
P_{ee}=\cos^2\xi\left[P_c^L\cos^2\omega+(1-P_c^L)\sin^2\omega+\sqrt{P_c(1-P_c^l)}
\sin 2\omega\cos\left(\frac{\Delta m^2_{21}L}{2E_{\nu}}\right)\right],
\end{equation}
while the pure vacuum result in the same region of the parameter space is
\begin{equation}
P_{ee}^{\rm vac}=\cos^4\xi\left[\cos^4\omega+\sin^4\omega+2\sin^2\omega\cos^2\omega
\cos\left(\frac{\Delta m^2_{21}L}{2E_{\nu}}\right)\right]+\sin^4\xi.
\end{equation}
In the limit $P_c^L=\cos^2\omega$, the difference  
\begin{equation}
P_{ee}-P_{ee}^{\rm vac}=\left(P_{ee}^{2\nu, \rm vac}\right)
\cos^2\xi\sin^2\xi-\sin^4\xi,
\end{equation}
where $P_{ee}^{2\nu, \rm vac}$ is the electron neutrino survival probability 
in the  two-flavour case with $\Delta m^2=\Delta m^2_{21}$ and vacuum mixing angle
$\omega$. This difference 
vanishes at $\sin^2\xi=0$, and $\sin^2\xi=\frac{P_{ee}^{2\nu, \rm vac}}{1+
P_{ee}^{2\nu, \rm vac}}$, (which is between 0 and 0.5). Furthermore, it is 
a convex function of $\sin^2\xi$, which means that $P_{ee}$ is {\sl larger}
than the pure vacuum case for values of $\sin^2\xi< \frac{P_{ee}^{2\nu, \rm vac}}{1+
P_{ee}^{2\nu, \rm vac}}$. Away from the limit $P_c^L=\cos^2\omega$, keeping in mind that
the oscillatory terms average out, $P_{ee}$ is still larger than the pure vacuum 
case if $\cos^2\omega>\sin^2\omega$ since 
$P_c^L\leq\cos^2\omega$, as one can easily verify. 

Also, in the limit $P_2^H=1$, $\sin^2\theta=1/2$,
\begin{equation}
P_{\mu\mu}=\frac{1}{2}\left[(1-P_c^L)U_{\mu1}^2+P_c^LU_{\mu2}^2+
U_{\mu3}^2+2\sqrt{P_c^L
(1-P_c^L)}U_{\mu1}U_{\mu2}\cos\left(\frac{\Delta m^2{21}L}{2E_{\nu}}\right)\right].
\end{equation}
The same expression applies for $P_{\tau\tau}$ with $U_{\mu i}\rightarrow U_{\tau i}$.
This is a consequence of $\sin^2\theta=\cos^2\theta$. Furthermore, in the limit 
$\sin^2\xi\rightarrow 0$ (and for $\sin^2\theta=\cos^2\theta$), $U_{\mu i}=U_{\tau i}$,
which explains why $P_{\mu\mu}=P_{\tau\tau}$ for $\sin^2\xi\lesssim 10^{-2}$. At VAC
this equality remains for all values of $\sin^2\xi$. The reason for this is that,
at VAC, the expression simplifies tremendously and
$P_{\mu\mu}=P_{\tau\tau}=\frac{1}{4}\left(1+\cos^2\xi\right)$.
In the same region of the parameter space, the pure vacuum oscillation case 
yields $P_{\mu\mu}^{\rm vac}=P_{\tau\tau}^{\rm vac}=\frac{1}{2}\cos^4\xi-\cos^2\xi+1$.
Note that, in this region of the parameter space 
$P_{\mu\mu}^{\rm vac}\geq P_{\mu\mu}$, the inequality being saturated at $\cos^2\xi=1$.

The same result also applies (approximately) at SMA, since the oscillatory terms are
proportional to $\sqrt{P_c^L(1-P_c^L)}$ and $1-P_c^L$ is very small at SMA (see
Fig~\ref{1-pcl}). The equality $P_{\mu\mu}=P_{\tau\tau}$ is broken at larger values of 
$\sin^2\xi$ because $P_c^L\neq \cos^2\omega$ at SMA.  

It remains to discuss how $P_{\mu\mu}$ and $P_{\tau\tau}$ diverge from the
pure vacuum case at LMA and LOW. In the limit $P_c^L=\cos^2\omega$, and averaging
out the oscillatory terms,
\begin{equation}
\label{eq_diff}
P_{\mu\mu}-P_{\mu\mu}^{\rm vac}=\frac{\sin\xi}{2}\left[\sin\xi\left(U_{\mu3}^2-
(\cos^2\omega U_{\mu1}^2+\sin^2\omega U_{\mu2}^2)\right)-
\sin 2\omega(U_{\mu1}^2-U_{\mu2}^2)
\right].
\end{equation}

This difference goes to zero as $\sin^2\xi\rightarrow 0$. This is to be expected, since
in this limit the difference of $P_2^H$ and $\cos^2\xi$ disappears. For small values
of $\sin^2\xi$, the last term in Eq.~\ref{eq_diff} dominates, and, as discussed
before, $U_{\mu1}^2-U_{\mu2}^2=-0.5\cos 2\omega+O(\sin\xi)$. Therefore, 
$P_{\mu\mu}-P_{\mu\mu}^{\rm vac}>0$ ($<0$) for $\cos 2\omega>0$ ($<0$). 
The expression for $P_{\tau\tau}$ can be obtained from Eq.(\ref{eq_diff}) by replacing 
$U_{\mu i}\rightarrow U_{\tau i}$ and changing the sign of the last term. 
Therefore, since $U_{\tau1}^2-U_{\tau2}^2=-0.5\cos 2\omega+O(\sin\xi)$,
$P_{\tau\tau}-P_{\tau\tau}^{\rm vac}>0$ ($<0$) for $\cos 2\omega<0$ ($>0$). When the 
oscillatory terms do not average out, it is easy to verify explicitly that 
the behaviour of the oscillatory terms follows the behaviour of the average terms, 
discussed above, and the inequalities obtained above still apply.
\begin{figure} [t]
\centerline{
  \psfig{file=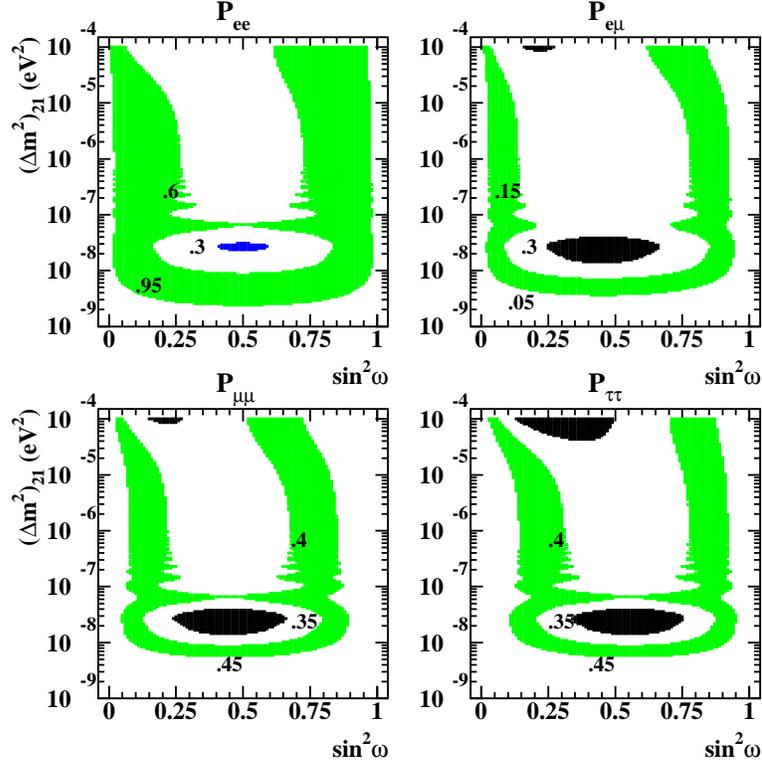,width=0.75\textwidth}
}
\caption{Constant $P_{\alpha\beta}$ contours in the 
($\Delta m^2_{21}\times\sin^2\omega$)-plane, at ATM (see text), for 
$\Delta m^2_{31}<0$.}
\label{sso_min}
\end{figure}

The situation, however, changes, when $P_c^L\neq \cos^2\omega$, {\it i.e.,}\/ when
matter effects due to the ``M-L'' system are relevant. In this region, a behaviour
similar to the one observed in the ``normal'' hierarchy case is expected, since
$\Delta m^2_{21}>0$.  
Fig.~\ref{sso_min} depicts constant $P_{\alpha\beta}$ contours in the 
($\Delta m^2_{21}\times\sin^2\omega$)-plane. One should be able to see upon close 
inspection that the region $P_{ee}<30\%$ is smaller in Fig.~\ref{sso_min} than the 
same region in the pure vacuum oscillation case, Fig~\ref{dm21_vacuum}. Also, the 
constant $P_{\mu\mu}$ ($P_{\tau\tau}$) contours are shifted to larger (smaller) 
values of $\sin^2\omega$. The other prominent (and expected, as mentioned above)
feature is the distortion of the contours at large values of $\Delta m^2_{21}$. 
This behaviour is similar to the one observed in Fig.~\ref{dm21_std}.


I conclude this subsection with a comment on antineutrinos. As discussed previously,
$P_{\bar{\alpha}\bar{\beta}}(\Delta m^2_{21},\Delta m^2_{31})
=P_{\alpha\beta}(-\Delta m^2_{21},-\Delta m^2_{31})$, such that the ``normal''
hierarchies yield ``inverted'' hierarchy results for antineutrinos, and vice-verse. 
One cannot, however, apply Fig.~\ref{dm21_std} and Fig.~\ref{sso_min} for the 
antineutrinos because both $\Delta m^2_{ij}$ have to change sign, not just 
$\Delta m^2_{31}$. Qualitatively, however, it is possible to understand the constant
$P_{\bar{\alpha}\bar{\beta}}$ contours by examining figures Fig.~\ref{dm21_std} and 
Fig.~\ref{sso_min} reflected in a mirror positioned at $\sin^2\omega=0.5$, meaning
that $P_{\bar{\alpha}\bar{\beta}}(\sin^\omega,\Delta m^2_{31})\simeq
P_{\alpha\beta}(\cos^2\omega,-\Delta m^2_{31})$. The equality is not complete
because one is also required to exchange $\theta\rightarrow \pi-\theta$, as mentioned
earlier. 

\subsection{Higher Neutrino Energies}

As the average neutrino energy increases, the values of $P_{\alpha\beta}$
start to resemble more the pure vacuum case. This is easy to see from 
Figs.~\ref{p3h} and \ref{1-pcl}. Any deviation of $1-P_c^L$ from $\sin^2\omega$ 
goes away even at LMA for $E_{\nu}\simeq 50$~GeV, while ``H-M'' effects remain important
up to $E_{\nu}\simeq 1$~TeV, even though quantitatively the effect decreases noticeably.
This can be illustrated by the value of $P_3^H$ at ATM, for example, which drops 
from 0.87 for  energies which range from 1 to 5~GeV 
(see the previous subsections) to 0.058, for 
energies which range from 100 to 110~GeV.

Furthermore, all $L^{\rm osc}_{ij}$ increase as the energy increases, 
for fixed values of $\Delta m^2_{ij}$. Therefore, LOW becomes indistinguishable from VAC
at $E_{\nu}\simeq 100$~GeV. For $O$(TeV) 
neutrinos the sensitivity to $\Delta m^2_{21}$ remains only for its highest allowed 
values, while one should start worrying about nontrivial oscillatory effects due to 
$L_{31}^{\rm osc}$.  

The case of higher energy neutrinos contains a more serious complications: neutrino
absorption inside the Sun. As the neutrino energy increases, one has to start worrying
about the fact that absorptive neutrino interactions can take place. According to
\cite{absorption}, for neutrinos produced in the Sun's core, absorption becomes 
important for $E_{\nu}\gtrsim 200$~GeV. In this case, $\nu_e$ and $\nu_\mu$ interact 
with nuclear matter and produce electrons and muons, respectively. 
The former are capture and 
``lost'' inside the Sun, while the latter stop before
decaying into low energy neutrinos. The case of $\nu_{\tau}$-Sun interactions is more
interesting, because the $\tau$-leptons produced via charged current interactions 
decay before ``stopping'', yielding $\nu_{\tau}$'s with slightly reduced energies.
Therefore, it is possible to get a flux of very high energy initial state
$\tau$-neutrinos but not
muon or electron-type neutrinos. 
Such effects have been studied for high energy galactic
neutrinos traversing the Earth \cite{absorption_earth}.    

The effect of neutrino oscillations inside the Sun in the presence of nonnegligible
neutrino absorption is certainly of great interest but is beyond the scope of this 
paper.

\section{Conclusions}

The oscillation probability of $O$(GeV) neutrinos of all flavours produced in the
Sun's core has been computed, including matter effects, which are, in general,
nontrivial.

In particular, it was shown that, unlike the two-flavour oscillation case, in the
three-flavour case the probability of a neutrino produced in the flavour eigenstate
$\alpha$ to be detected as a flavour eigenstate $\beta$ ($P_{\alpha\beta}$) is 
(in general) different from $P_{\beta\alpha}$, even if the $CP$-violating phase
of the MNS matrix vanishes. This is, of course, expected since Sun--neutrino 
interactions explicitly break $T$-invariance. Indeed, it is the case of two-flavour 
oscillations which is special, in the sense that the number of independent oscillation 
probabilities is too small because of unitarity. 

The results of a particular scan of the parameter space are presented in Sec.~4. 
In this case, special attention was paid to the regions of the parameter space
which are preferred by the current experimental situation.

It turns out that, in the case of a ``normal'' neutrino mass hierarchy,
it is possible to suppress $P_{ee}$ tremendously with respect to its
pure vacuum oscillation values, by a mechanism that is similar to the well known
MSW effect in the case of two-flavour oscillations: the parameters are such that 
electron-type neutrinos produced in the Sun's core exit the Sun (almost) as pure
mass eigenstates, and the $\nu_e$ component of this eigenstate is small. Both 
$P_{\mu\mu}$ and $P_{\tau\tau}$ can be significantly suppressed, and the constant 
$P_{\mu\mu}$ and $P_{\tau\tau}$ contours as a function of the ``solar'' angle and
the smaller mass-squared differences are nontrivial. One important feature is
that when $P_{\mu\mu}$ is significantly suppressed, $P_{\tau\tau}$ is not, and 
vice-versa. One consequence of this is that, for some regions of the parameter space,
it is possible to have an enhancement of $\nu_{\tau}$'s detected in the Earth with
respect to the number of $\nu_{\mu}$'s (or vice-versa). This may have important 
implications for solar WIMP annihilation searches at neutrino telescopes, and will 
be studied in another oportunity. It is important to note that the effect of 
neutrino oscillations on the expected event rate at neutrino telescopes 
will depend on the expected production rate of individual neutrino
species inside the Sun, which is, of course, model dependent.  

In the case of an ``inverted'' mass hierarchy, the situation is very similar to the 
pure vacuum case, and no particular suppression of any $P_{\alpha\alpha}$ is possible. 
Indeed, for a large region of the parameter space $P_{ee}$ is in fact enhanced, 
a feature which is also observed in the two-flavour case \cite{earth_matter}.    

The case of higher energy neutrinos was very briefly discussed, and the crucial point
is to note that, for neutrino energies above a few hundred GeV, the absorption of
neutrinos by the Sun becomes important. The study of absorption effects is
beyond the scope of this paper.

Finally, it is important to reemphasise that the values of $P_{\alpha\beta}$ computed
here are to be understood as if they were evaluated at the Earth's surface. No 
Earth-matter effects have been included. It is possible that Earth-matter effects 
are important, especially the ones related to $\Delta m^2_{31}$, in the advent that
$U_{e3}^2\equiv\sin^2\xi$ turns out to be ``large.'' 

\section*{Acknowledgements} 

I would like to thank John Ellis for suggesting the study of GeV solar 
neutrinos, and for
many useful discussions and comments on the manuscript. I also thank Amol 
Dighe and Hitoshi Murayama for enlightening discussions and for carefully 
reading this manuscript and providing useful comments.

\end{document}